\documentclass[11pt]{article}

\usepackage[margin=1in]{geometry}
\usepackage[T1]{fontenc}
\usepackage{lmodern}
\usepackage{microtype}
\usepackage{setspace}
\usepackage{graphicx}
\usepackage{xcolor}
\usepackage{authblk}
\usepackage{titlesec}
\usepackage{abstract}
\usepackage{csquotes}
\usepackage[font=small]{caption}
\usepackage{soul}%
\usepackage{xcolor,colortbl} %
\usepackage{changepage,threeparttable} %
\usepackage{microtype}
\usepackage{booktabs, multirow} %
\usepackage{float} %
\usepackage{amsmath,amssymb}
\usepackage{amsmath,amssymb}
\usepackage{booktabs}
\usepackage{longtable}
\usepackage{array}
\usepackage{hyperref}
\usepackage{tabularx}
\usepackage{makecell}

\definecolor{paperblack}{HTML}{1F1F1F}
\definecolor{mutedblue}{HTML}{5B8FB9}

\usepackage[
backend=biber,
style=authoryear-comp,
maxcitenames=2,
mincitenames=1,
maxbibnames=2,
giveninits=true,
uniquelist=false,
uniquename=false,
date=year,
doi=true,
url=false,
isbn=false,
natbib=true,
eprint=false,
]{biblatex}

\DeclareSourcemap{
  \maps[datatype=bibtex]{
    \map[overwrite]{
      \step[fieldsource=entrykey,
            match=\regexp{^golombek_matthew_spacex_2021$},
            final]
      \step[fieldset=author,
            fieldvalue={Golombek, Matthew and Williams, Nathan and Wooster, Paul and McEwen, Alfred S. and Putzig, Nathaniel E. and Bramson, Ali M. and Head, James and Heldmann, Jennifer and Marinova, Margarita and Beaty, David}]
    }
  }
}

\AtEveryBibitem{%
  \clearfield{url}%
  \clearfield{urldate}%
  \clearfield{eprint}%
  \clearfield{eprinttype}%
  \clearfield{eprintclass}%
  \clearfield{note}%
  \clearfield{eventtitle}%
}

\DeclareCiteCommand{\bracketcite}[\mkbibbrackets]
{\usebibmacro{prenote}}
{\usebibmacro{citeindex}%
\usebibmacro{cite}}
{\multicitedelim}
{\usebibmacro{postnote}}

\let\citep\bracketcite
\let\citet\textcite

\usepackage{hyperref}

\hypersetup{
colorlinks=true,
linkcolor=paperblack,
citecolor=mutedblue,
urlcolor=mutedblue,
pdftitle={Doubling Sunlight for A Human Base on Mars},
pdfauthor={Ari Essunfeld, Yuji Takubo, Adrian Dumitrescu, Fabrizio Pisani, Casey Handmer, Erika DeBenedictis, Edwin Kite}
}

\titleformat{\section}
{\large\sffamily\bfseries\color{paperblack}}
{\thesection}
{0.75em}
{}

\titleformat{\subsection}
{\normalsize\sffamily\bfseries\color{paperblack}}
{\thesubsection}
{0.75em}
{}

\titleformat{\subsubsection}
{\normalsize\sffamily\bfseries\color{paperblack}}
{\thesubsubsection}
{0.75em}
{}

\titlespacing*{\section}{0pt}{1.5em}{0.6em}
\titlespacing*{\subsection}{0pt}{1.1em}{0.4em}

\title{\vspace{-1.5em}\sffamily\bfseries\Large Doubling Sunlight for a Human Mars Base \\ With Orbiting Solar Reflectors}

\author[1,2]{Ari Essunfeld (ari.essunfeld@princeton.edu)}
\author[3]{Yuji Takubo}
\author[1]{Adrian Dumitrescu}
\author[1]{Alexandre Kling}
\author[1]{Fabrizio Pisani}
\author[4]{Casey Handmer}
\author[1,5]{Edwin Kite}

\affil[1]{Astera Institute, Emeryville, CA}
\affil[2]{Department of Mechanical and Aerospace Engineering, Princeton University, Princeton, NJ}
\affil[3]{Department of Aeronautics and Astronautics, Stanford University, Stanford, CA}
\affil[4]{Terraform Industries, Burbank, CA}
\affil[5]{Department of Geophysical Sciences, University of Chicago, IL}

\date{\vspace{-1em}September 2026\vspace{-1em}}

\newcommand{\nhat}{\hat{\mathbf n}}
\newcommand{\shat}{\hat{\mathbf s}}
\newcommand{\hhat}{\hat{\mathbf h}}
\newcommand{\eA}{\hat{\mathbf e}_A}
\newcommand{\eB}{\hat{\mathbf e}_B}
\newcommand{\that}{\hat{\mathbf t}}
\newcommand{\vhat}{\hat{\mathbf v}}
\newcommand{\zhat}{\hat{\mathbf z}}

\begin{document}

\maketitle

\vspace{-1.5em}

\begin{abstract}

The Sun's faintness at Mars' orbit makes producing energy, melting ice, and staying warm more~difficult. 
Orbiting solar reflectors (OSRs) can augment sunlight at Mars, but~the area of OSRs needed to~double sunlight at a~Mars base is not known.
Here, we~analyze Sun-synchronous Mars orbits to find the OSR area that doubles insolation to~a~Mars base. 
We~show that the reflectors can deliver sunlight and maintain a~stable orbit via~attitude control and solar-sail propulsion, with no~propellant.
We also show that these Mars orbits can be~reached via solar sailing from low Earth orbit, reducing delivery cost. 
Doubling sunlight is viable with an orbiting solar reflector areal density of $20~{\rm g/m^2}$, a~7-fold improvement over flight-proven solar sails. 
However, designing a~spacecraft with low areal density, agile maneuverability (for accurate pointing), high~tension in the sail membrane (to smooth wrinkles), and mass-manufacturability (to enable a~constellation of reflectors) would be challenging.
Still lower areal density of $<5~{\rm g/m^2}$ would be needed for more ambitious applications, such as sublimating the CO$_2$ ice at Mars' south pole to~aid~in terraforming.

\end{abstract}

\section{Introduction}
\label{sec:introduction}
Mars receives much less sunlight than does the Earth. 
To restore Mars' past habitability [\cite{mangold_perseverance_2021}], warming would be a necessary prerequisite  [\cite{mckay_physics_2001}, \cite{debenedictis_case_2025}]. 
Proposed warming methods either (1)~strengthen the~greenhouse effect [e.g., \cite{mckay_making_1991},  \cite{marinova_radiative-convective_2005}, \cite{wordsworth_enabling_2019}, \cite{ansari_feasibility_2024}, \cite{richardson_atmospheric_2026},
\cite{braude_modelling_2026}, \cite{turyshev_terraforming_2026}], or (2)~direct more sunlight to Mars. 

Taking the second approach, we ask: 
what is the smallest combined reflector area for which orbiting solar reflectors (OSRs) give major benefits to a Mars base? In principle, OSRs could help supply power, warmth, and meltwater for a Mars base. 
A minimal system could be a technological stepping-stone toward bigger systems (e.g., \citealt{handmer_how_2024,zubrin_technological_1993}).
Previous work considered OSRs to warm the whole of Mars [\cite{mcinnes_non-keplerian_2002}, \cite{mcinnes_mars_2009}].
\citet{salazar_sun-synchronous_2019} studied Sun-synchronous frozen-eccentricity orbits for Mars-warming. 
OSRs in Earth orbit might support solar farms on Earth's surface [e.g., \cite{viale_reference_2023,celik_analytical_2022, celik_generic_2023, celik_constellation_2024}; and Reflect Orbital\footnote{\url{https://www.reflectorbital.com/}}].
However, neither orbital stability nor energy efficiency have been considered for a Mars-base application.
Given Mars' high eccentricity ($e\approx0.09$) and the correspondingly variable solar radiation pressure, it is reasonable to question if a low-altitude Mars-orbiting solar reflector could maintain orbital stability.

As a proxy for a major benefit, we use local doubling of average insolation. 
Several applications motivate this threshold: 
sustaining a minimal human~base on Mars would likely require at least MWe power to make propellant (scaling from \citet{hinterman_multi-objective_2022} and \citet{gentgen_bart_2022} to Starship-class vehicles), and even more for a bigger base.
If solar energy supplies most of the~power, Mars' distance from the Sun doubles the panel area needed, relative to Earth.
When the marginal cost of boosting existing solar panels using light from OSRs is less than for adding more solar panels, OSRs are favored. 
Doubling~sunlight also warms peak daytime temperatures above freezing for a base at 40$^\circ$N for the~whole year
(Section~\ref{res:microclimate-modeling-results}).

\begin{figure}[h]
    \centering
    \includegraphics[width=0.8\textwidth]{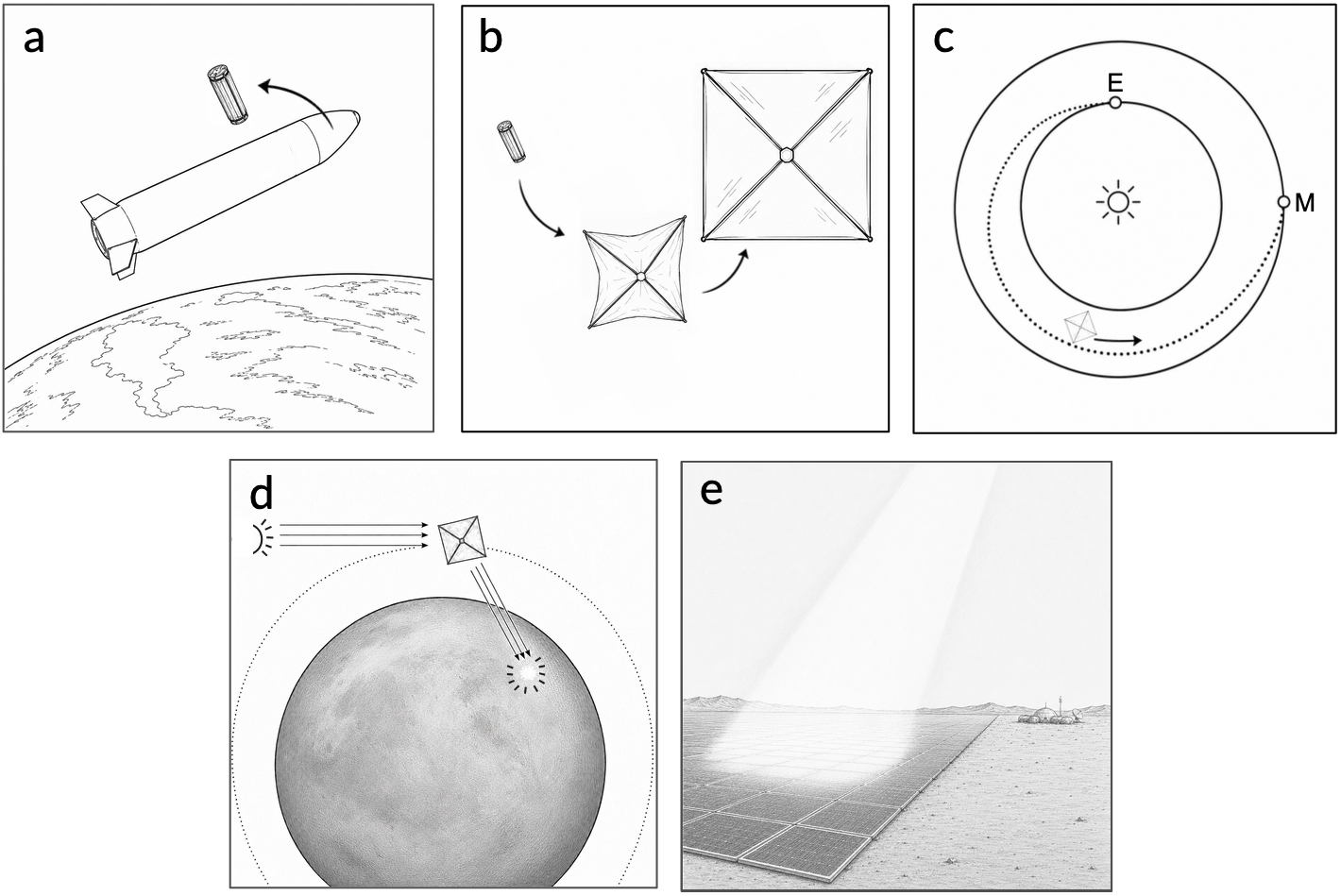}
    \caption{
    Concept of operations.
    \textbf{(a)}: Orbiting solar reflector (OSR) delivered to 800~km altitude Earth orbit.
    \textbf{(b)}: Reflector deploys in Earth orbit.
    \textbf{(c)}: Reflector escapes from Earth, sails to Mars, and enters Mars orbit.
    \textbf{(d)}: Reflector in Mars orbit supplies sunlight to a human~base.
    \textbf{(e)}: Solar array near human~base illuminated by OSR. In practice, the beam would overspill the array (Appendix~\ref{app:reflected-spot}).}
    \label{fig:conops}
\end{figure}

Implementing Mars OSRs as solar-sail spacecraft and flying them from Earth to Mars under their own solar radiation pressure propulsion [\cite{hughes_realistic_2005}] takes advantage of the relative ease (and wider range of launch vehicles) for reaching low Earth orbit (LEO) than for reaching Mars orbit.
Solar sails were flight-demonstrated by~JAXA's \textit{IKAROS} [\cite{tsuda_achievement_2013}] and the Planetary Society's 140~g/m$^2$ \textit{LightSail 2}  [\cite{spencer_lightsail2_2021}]. 
For Mars, putting a capability in orbit can be much easier than putting that capability on Mars's surface. 
For example, no bulky entry, descent and landing (EDL) system is required. 
This is in contrast to Earth, where the cost of access to orbit is much larger than the cost of ground transport. 

Here we assess solar-sail OSRs for doubling insolation at a~base on Mars (Figure \ref{fig:conops}).
The~specific contributions of this paper are as follows:
\begin{enumerate}
    \item We show that a solar-sail OSR could reach low Mars orbit (LMO) from low Earth orbit (LEO) via solar sailing (Figs.~\ref{fig:get-oriented}--\ref{fig:transfer}).
    \item We identify four candidate families of continuously sunlit Sun-synchronous orbits that OSRs could occupy to illuminate a~base at 40$^\circ$N, 200$^\circ$E (Fig.~\ref{fig:feasible-LTANs}).
    \item We develop a steering law that enables an OSR, subject to realistic slew constraints, to deliver sunlight and maintain stability in LMO; we demonstrate the steering law with 1- and 3-year simulations  (Figs.~\ref{fig:nonilluminating-orbit}--\ref{fig:illuminating-orbit}).
    \item We present a constellation concept that doubles insolation at a human~base using $\approx1{,}400$~km$^2$ of reflector surface area in orbit (Figs.~\ref{fig:densify-K12}--\ref{fig:multishell-power-to-base}). 
    One configuration for this constellation has $\approx10^5$ square-geometry $120~{\rm m}\times120~{\rm m}$ solar-sail OSRs; an energy-equivalent configuration with fewer, individually larger reflectors may be~needed for operational collision safety (Appendix~\ref{app:constellation-sizing}).
    For 20 g/m$^2$ areal density, deployment would require 100--200 Starship-class launches to LEO.
\end{enumerate}

We also show energy delivered as a function of time-of-day, benefits for battery sizing, and the surface temperatures resulting from orbiting-solar-reflector heating (Figs.~\ref{fig:multishell-power-to-base}--\ref{fig:temp-vs-Ls}).


\section{Methods}
\label{sec:methods}

We simulate
(a) flight of a solar-sail OSR from LEO to LMO;
(b) long-term stability of an OSR that delivers sunlight to a human base while in Mars orbit; and
(c) energy supply to the base.
All~software is open-source on GitHub\footnote{\url{https://github.com/ariessunfeld/mars-osr}} and archived on Zenodo\footnote{\url{https://doi.org/10.5281/zenodo.22168117}}.

The Methods are organized as follows: Section~\ref{methods:sail-model} describes our sail model.
Section~\ref{methods:reference-frames} describes the reference frames used for Earth escape, interplanetary transfer, Mars capture, and low Mars orbit.
Section~\ref{methods:forces-and-perturbations} states the forces and perturbations modeled.
Section \ref{methods:LEO-to-LMO} describes how we find LEO $\to$ LMO solar sail trajectories and the control laws and optimizations used for each part of the~trajectory segment.
Section \ref{methods:delivering-sunlight} describes how a reflector maintains orbital stability while delivering sunlight to a Mars base.
Section \ref{methods:scaling-up} describes how we calculate the energy delivered by many reflectors and scale those calculations to a constellation of orbiting solar reflectors.

\subsection{Sail model}
\label{methods:sail-model}

We model the~orbiting solar reflector as a flat, rigid, square, imperfect reflector. 
Solar radiation pressure acceleration $\mathbf a_{s}$ is calculated using the~model of \citet{mcinnes_solar_1999}'s Eq.~2.57, with solar pressure $P(r_\odot)=L_\odot/(4\pi c\, r_\odot^2)$ evaluated at the~sail's heliocentric distance $r_\odot$. $c$ is the~speed of light.
We use solar luminosity $L_\odot = 3.828 \times 10^{26}~{\rm W}$ [\cite{prsa_nominal_2016}].
We use the~optical coefficients of an aluminized square sail [\cite{mcinnes_solar_1999}'s Table~2.1] and values for sail area $A$ and mass $m$ consistent with its assumed areal density~$\sigma = m/A$.
Solar radiation pressure is set to zero during eclipses, which are determined using a conical shadow model. We~do not model reflectivity degradation over time.

Rather than allowing instant sail reorientation, we~model the~sail as having finite, achievable agility.
Thus, we~bound the~angular-velocity vector of the~sail normal, $\boldsymbol\omega = \nhat\times \dot{\nhat}$, and the~angular-acceleration vector of the~sail normal,
$\dot{\boldsymbol\omega} = \nhat\times\ddot{\nhat}$:
\begin{equation}
    \lVert\boldsymbol\omega\rVert \le \omega_{\max}=0.3~\mathrm{deg/s},
    \qquad
    \lVert\dot{\boldsymbol\omega}\rVert \le \dot\omega_{\max}=3\times10^{-3}~\mathrm{deg/s^2}
    \label{eq:esc-slew}
\end{equation}

for all trajectory segments.
These values are similar to those used in \citet{viale_attitude_2023}.

\subsection{Reference frames and coordinate systems}
\label{methods:reference-frames}

\begin{figure}[h]
    \centering
    \includegraphics[width=0.7\textwidth]{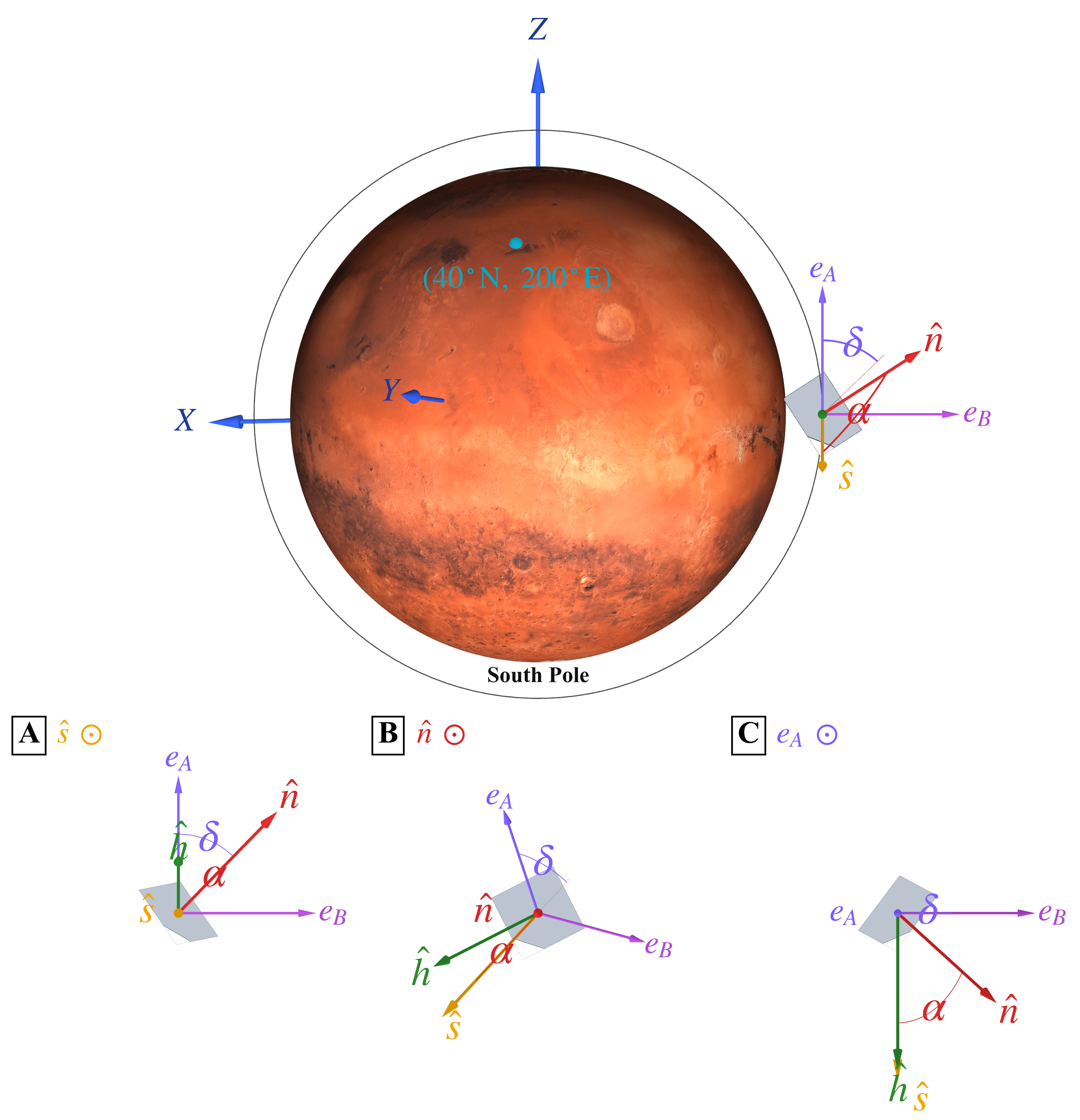}
    \caption{Low Mars Orbit. The Sun-referenced cone/clock frame (see Section \ref{methods:reference-frames} for definitions of vectors and~angles) in which the sail normal $\nhat$ is written. Orientations A--C show the same diagram, but with different vectors facing the viewer, as a visual aid. \textbf{A}: The Sun-line $\shat$ faces the viewer. \textbf{B:} The sail-normal $\nhat$ faces the viewer. \textbf{C:} $\eA$ faces the~viewer.}
    \label{fig:get-oriented}
\end{figure}

Reference frames are as follows.
For Earth escape, interplanetary transfer, and Mars capture, we integrate each trajectory segment in its own body-centered, inertial J2000 frame: Earth-centered for Earth escape, Sun-centered for interplanetary transfer, and Mars-centered for~Mars capture.
We patch the segments at the Earth and Mars Hill~spheres.
In each segment, as in low Mars orbit (below), we reference the sail attitude to the Sun-line $\shat$: the commanded sail normal $\nhat$ is written as a tilt off $\shat$.
The secondary axes about which $\nhat$ is tilted vary between segments: for Earth escape and Mars capture, $\nhat$ is tilted toward the velocity direction within the $(\shat,\vhat)$ plane (see Eq.~\ref{eq:esc-normal}); for interplanetary transfer, it is tilted within and out of the orbit plane using the heliocentric orbit normal $\hhat$; and for low Mars orbit, it is tilted using the orbit-normal-referenced clock-frame basis $(\eA,\eB)$ defined below.

In Mars orbit, we use the \citet{mcinnes_solar_1999} Sun-referenced
cone/clock frame (Figure \ref{fig:get-oriented}) in which the sail normal $\nhat$ is written as
\begin{equation}
  \nhat = \cos\alpha\,\shat
        + \sin\alpha\,\bigl(\cos\delta\,\eA + \sin\delta\,\eB\bigr),
  \label{eq:normal}
\end{equation}
where $\shat$ is the Sun-line unit vector (sail $\to$ Sun), $\alpha$ is the cone angle between $\nhat$ and the Sun-line, and $\delta$ is the clock angle (azimuth of $\nhat$ about the Sun-line). The clock-frame basis is built from the orbit-normal unit vector $\hhat$:
\begin{equation}
  \eA = \frac{\hhat - (\hhat \cdot \shat)\,\shat}
             {\bigl\lVert \hhat - (\hhat \cdot \shat)\,\shat \bigr\rVert},
  \qquad
  \eB = \eA \times \shat .
  \label{eq:basis}
\end{equation}
The vector $\eA$ is the orbit normal $\hhat$ projected into the plane perpendicular to the Sun-line (and then normalized). The triad $(\shat,\eA,\eB)$ is right-handed. Because $\nhat\cdot\shat = \cos\alpha$, imposing $\alpha\in[0,\pi/2]$ ensures that the non-reflective face of the sail is never sunlit.

\subsection{Forces, perturbations, ephemerides, and kernels}
\label{methods:forces-and-perturbations}

We integrate the~equations of motion of each trajectory segment in the~corresponding body-centered inertial J2000 frame (see Section~\ref{methods:reference-frames}) using data from the~Spacecraft, Planet, Instrument, Camera-matrix, Events (SPICE) Toolkit \citep{acton_look_2018} via SpiceyPy \citep{annex_spiceypy_2020}.
We load the~Jet Propulsion Laboratory (JPL) DE440 planetary ephemerides \citep{park_jpl_2021} (\texttt{de440.bsp}) for the~Sun, Earth, Moon, and Mars; 
the Navigation and Ancillary Information Facility (NAIF) Mars satellite ephemeris (\texttt{mar099.bsp}) for the~positions of Mars and its moons; 
and the~planetary-constants kernel (\texttt{pck00011.tpc}) for body radii and spin-pole orientations, among other kernels for gravitational parameters and leap-second corrections. 
Physical constants come from these kernels wherever possible.
Gravitational parameters $\mu$ are read from \texttt{gm\_de440.tpc}; the~value $\mu_\oplus$ in Eq.~\ref{eq:esc-energy} is this DE440 value.
For Earth's oblateness we use the~tide-free $J_2 = 1.0826267\times10^{-3}$ of the~Earth Gravitational Model 2008 [\cite{montenbruck_satellite_2000}].
For Mars we use the~\texttt{MRO120F} gravity field \citep{konopliv_detection_2020} to degree and order six.

We model the~dominant forces for each trajectory segment (Table~\ref{tab:forces-reference-frames}).
During Earth escape, the~sail begins in low Earth orbit, so we include Earth's oblateness, third-body gravity from the~Sun and Moon, solar radiation pressure, and atmospheric drag.
During interplanetary transfer, solar gravity and solar radiation pressure dominate the~sail's dynamics; we retain perturbations from Earth, the~Moon, and Mars.
During Mars capture, we model Mars as a point mass plus $J_2$, with the~Sun as the~sole third body.
In low Mars orbit, we raise the~Mars field to degree and order six and include Phobos' and Deimos' gravity.
Solar radiation pressure is computed using the~non-ideal optical flat-sail model of \citet{mcinnes_solar_1999} (Section~\ref{methods:sail-model}) in every segment.
We do not model pressure from light reflected from any celestial bodies, nor do we model thermal radiation pressure from any celestial bodies.

\begin{table}[!htp]
\centering
\caption{Forces, reference frames, and integrators used for each trajectory segment. See text for details.}
\label{tab:forces-reference-frames}
\resizebox{\textwidth}{!}{%
    \begin{tabular}{lllll}
    \toprule
    & Earth escape & Interplanetary transfer & Mars capture & Low Mars orbit \\
    \midrule
    Central gravity & Earth point-mass & Sun point-mass & Mars point-mass & Mars point-mass \\
    Non-spherical gravity & Earth $J_2$ & none & Mars $J_2$ & Mars (degree \& order 6) \\
    Third bodies & Sun + Moon & Earth, Moon, Mars & Sun & Sun, Phobos, Deimos \\
    Atmospheric drag & Harris--Priester \citep{montenbruck_satellite_2000} & none & none & none \\
    SRP + sail & \citet{mcinnes_solar_1999} non-ideal optical flat sail & $\leftarrow$ same & $\leftarrow$ same & $\leftarrow$ same \\
    Eclipse & binary umbra & none & binary umbra & binary umbra \\
    Integration frame & Earth-centered J2000 & Heliocentric J2000 & Mars-centered J2000 & Mars-centered J2000 \\
    Integration algorithm & \texttt{RK4} & \texttt{RK4} & \texttt{RK4} & \texttt{DOP853} \\
    \bottomrule
    \end{tabular}%
}
\par\vspace{3pt}{\footnotesize\raggedright
The Harris--Priester atmospheric-density model is tabulated in \citet{montenbruck_satellite_2000}'s \S3.5.\par}
\end{table}

\subsection{Solar sailing from LEO to LMO}
\label{methods:LEO-to-LMO}

Deploying a solar sail in low Earth orbit (LEO) and flying it via solar radiation pressure to low Mars orbit takes advantage of the relative ease (and wider range of launch vehicles) for reaching low Earth orbit (LEO) than for reaching Mars orbit.

Here we show that a $\sigma = 18$ g/m$^2$ (cf. \citealt{viale_reference_2023}), realistically slew-limited (Eq.~\ref{eq:esc-slew}) solar sail could do this. To fly from LEO to LMO, the sail must escape Earth (Section \ref{methods:earth-escape}), navigate to Mars (Section \ref{methods:interplanetary-transfer}), and be captured by Mars (Section \ref{methods:mars-capture}).
We show this with trajectories from LEO to LMO at a wide range of Earth–Mars phasings \citep{hughes_realistic_2005} (Figure \ref{fig:transfer}).
We solve escape and capture first, then the transfer that connects the handoff states, producing a consistent, continuous, flyable trajectory.

\subsubsection{Earth escape}
\label{methods:earth-escape}

To escape Earth, the~sail uses solar radiation pressure thrust to raise its specific orbital energy [\cite{prussing_solar_2000}, \cite{coverstone_technique_2003}]. For a geocentric state $(\mathbf r, \mathbf v)$, the~specific two-body energy is

\begin{equation}
    \varepsilon \; = \; \tfrac12\,\lVert \mathbf v \rVert^2 \; - \; \frac{\mu_\oplus}{\lVert \mathbf r \rVert},
    \label{eq:esc-energy}
\end{equation}
where $\mu_\oplus$ is Earth's gravitational parameter ($\mu_\oplus\,\approx\,3.986\times10^5$ km$^3$/s$^2$ [\cite{park_jpl_2021}]).

Maximizing orbital-energy gain tends to increase eccentricity, which can cause perigee to collapse [\cite{macdonald_realistic_2005}].
We use a greedy control law (described below) and enforce perigee safety through a minimum-altitude floor (Eq.~\ref{eq:esc-floor}).
Any trajectory that reaches the~floor is terminated and recorded as a failed escape.

\textit{Steering law}:
\label{methods:LEO_to_LMO:earth_escape:steering_law}
Under perturbing acceleration $\mathbf a$, the~specific orbital energy's rate of change is given by $\dot\varepsilon = \mathbf a\cdot\mathbf v$ (Appendix~\ref{app:energy-rate-derivation}). To maximize energy gain, the~control law maximizes the~velocity-aligned component of the~sail acceleration. Rather than use the~full Eq.~\ref{eq:normal} cone/clock parameterization, we~parameterize the~sail normal $\nhat$ with a single cone angle $\alpha$ in the~plane spanned by the~sail$\to$Sun line $\shat$ and the~sail's velocity vector:
\begin{equation}
    \nhat(\alpha) \;=\; \cos\alpha\,\shat \;+\; \sin\alpha\,\that,
    \label{eq:esc-normal}
\end{equation}
where
\begin{equation}
    \that \;=\; 
    \frac{\vhat - (\vhat \cdot \shat)\, \shat}{ \lVert\vhat - (\vhat\cdot\shat)\,\shat\rVert},
    \qquad
    \that \perp \shat \ \text{in the~}(\shat, \vhat)\text{ plane},
    \qquad
    \that \cdot \vhat \ge 0
\end{equation}
is the~in-plane toward-velocity direction (i.e., $\that$ is $\vhat$ with its Sun-ward part removed). At each integration step, we choose \begin{equation}
  \alpha^\star \;=\; 
  \operatorname*{arg\,max}_{\lVert\alpha\rVert\le\alpha_{\mathrm{c}}}\;
  \left[
  \mathbf a_s \bigl(\nhat(\alpha)\bigr) 
  + 
  \mathbf a_d \bigl(\nhat(\alpha)\bigr)
  \right]
  \cdot \vhat,
  \label{eq:esc-argmax}
\end{equation}
where cone bound $\alpha_{\mathrm c}$ is $80^\circ$, $\mathbf a_s\bigl(\nhat(\alpha)\bigr)$ is the~solar radiation pressure acceleration expressed as a function of $\nhat(\alpha)$, and $\mathbf a_d\bigl(\nhat(\alpha)\bigr)$ is the~acceleration due to atmospheric drag (see below).
For~the~maximization in Eq.~\ref{eq:esc-argmax}, we use a coarse grid search followed by golden-section refinement.
If~no~permitted orientation provides a positive net velocity-aligned acceleration, the~sail is oriented to minimize the~energy-decreasing acceleration (e.g., feathered edge-on to its velocity when drag dominates, or edge-on to the~Sun when solar radiation pressure dominates).
We assume that the~reflector can survive these brief edge-on periods (e.g., using batteries).

\textit{Atmospheric drag}:
\label{methods:LEO_to_LMO:earth_escape:atmospheric_drag}
Since the~sail has low areal density and begins its escape at 800 km altitude, drag is non-negligible. We model drag as
\begin{equation}
    \mathbf a_{d} \;=\; -\tfrac12\, \rho\, C_{d}\, \frac{A_{\mathrm{proj}}}{m}\,
    \lVert\mathbf v_{\mathrm{rel}}\rVert\,\mathbf v_{\mathrm{rel}},
    \qquad
    A_{\mathrm{proj}} \;=\; A\,\lVert\nhat\cdot\hat{\mathbf v}_{\mathrm{rel}}\rVert,
    \label{eq:esc-drag}
\end{equation}
where $C_d=2.2$, $\mathbf v_{\mathrm{rel}}$ is the~velocity relative to the~co-rotating atmosphere (with unit vector $\hat{\mathbf v}_{\mathrm{rel}}=\mathbf v_{\mathrm{rel}}/\lVert\mathbf v_{\mathrm{rel}}\rVert$), $A_{\mathrm{proj}}$ is the~sail area projected onto the~oncoming flow, and $\rho$ is the~altitude-dependent atmospheric density at mean solar activity from the~Harris--Priester model [\cite{montenbruck_satellite_2000}'s \S3.5].
Periods of elevated solar activity may require higher initial orbits.
The~objective in Eq.~\ref{eq:esc-argmax} considers both $\mathbf a_s$ and $\mathbf a_d$ so that the~controller is aware of both drag and solar radiation pressure gain, and could, for instance, feather edge-on to the~flow when drag would dominate.
We do not model sail material degradation (e.g., due to atomic oxygen); instead, we use a conservative altitude floor (Eq.~\ref{eq:esc-floor}).


The equations of motion are advanced with fourth-order Runge--Kutta using a nominal time step constrained by a cap on the~osculating true-anomaly advance, $\Delta t \le (r^2/h)\Delta\nu_{\max}$, corresponding to the~$n=2$ true-anomaly member of the~generalized Sundman family, $dt=c r^n ds$ \citep{berry_generalized_2002}.
This cap concentrates evaluations near perigee.
We also impose an absolute step-size cap to~keep the~attitude tracker converged.
We use a circular, $800$~km altitude, dawn--dusk, polar ($i=90^\circ$) initial orbit.

\textit{Escape criteria and altitude floor}:
\label{methods:LEO_to_LMO:earth_escape:esc_criteria}
The sail escapes Earth when (a) its specific orbital energy (Eq.~\ref{eq:esc-energy}) is non-negative and (b) its geocentric distance reaches Earth's Hill radius:
\begin{equation}
  \varepsilon \ge 0 \quad\text{and}\quad \lVert\mathbf r\rVert \ge r_{\mathrm H} \quad \implies \quad \text{escaped.}
  \label{eq:esc-criterion}
\end{equation}
If the~sail's geocentric radius falls to 600~km altitude:
\begin{equation}
  \lVert\mathbf r\rVert \;\leq\; R_\oplus + 600~\mathrm{km} \quad \implies \quad \text{non-escape.}
  \label{eq:esc-floor}
\end{equation}
We refer to the~geocentric position, velocity, and epoch when Eq.~\ref{eq:esc-criterion} is first satisfied as the~\textit{escape handoff state}, $X_{\mathrm{esc}}$:
\begin{equation}
    X_{\mathrm{esc}} \;=\; (\mathbf r_{\rm esc}, \, \mathbf v_{\rm esc}, \, t_{\mathrm{esc}})
    \label{eq:esc-state}
\end{equation}

\subsubsection{Mars capture}
\label{methods:mars-capture}

To reach low Mars orbit, the sail must be gravitationally captured by Mars [e.g., \cite{topputo_earth--mars_2015}]. 
We use solar radiation pressure to steer the sail into a capture spiral.
We design this capture spiral by running the Earth-escape procedure (Section~\ref{methods:earth-escape}) backward in time, starting from low Mars orbit.
This is possible due to the time-symmetry of the drag-free equations of motion used for~Mars capture.

We initialize the sail in a low Mars orbit at arrival epoch $t_{\mathrm a}$.
We use a circular, Sun-synchronous orbit at one of the target altitudes (see Section~\ref{res:families-of-orbits-for-MMaaS}) with initial LTAN of $L_0=18$~h and initial mean anomaly of $M_0 = 0^\circ$.
We then reverse the sail's velocity vector ($\mathbf{v}\,\to\,-\mathbf{v}$) and propagate it with the~same energy-maximizing steering law (Eq.~\ref{eq:esc-normal}) and integrator used for Earth escape. 
However, we step the ephemeris clock backwards, so that after a propagation interval $t$ the ephemeris time is $t_{\mathrm a}-t$;
and we omit atmospheric drag, since we treat Mars's atmosphere as negligible at~altitudes $> 300~{\rm km}$.
Under this backwards clock, the steering law spirals the sail outward, raising its Mars-relative specific energy until $\varepsilon \ge 0$ and its areocentric distance reaches Mars's Hill radius $r_{\mathrm H}$ (the same criterion as Eq.~\ref{eq:esc-criterion}).

Run forward in time, this backward-escape trajectory is a capture spiral that begins at Mars's Hill~sphere and ends in the desired low Mars orbit.
Because time reversal reverses velocity, the~forward~capture begins with the negated Hill-sphere velocity.
We define the \textit{capture handoff state} as
\begin{equation}
    X_{\mathrm{cap}} \;=\; (\mathbf r_{\rm cap}, \, -\mathbf v_{\rm cap}, \, t_{\mathrm{cap}}),
    \label{eq:cap-state}
\end{equation}
where $(\mathbf r, \mathbf v)$ is the areocentric state at which the backwards-escape criteria are first satisfied, $t_{\mathrm{cap}} = t_{\mathrm a} - T_{\mathrm{cap}}$ is the corresponding epoch, and $T_{\mathrm{cap}}$ the capture duration.
(While the escape handoff $X_{\mathrm{esc}}$ (Eq.~\ref{eq:esc-state}) is the \textit{end} of a forward-time \textit{escape}, $X_{\mathrm{cap}}$ (Eq.~\ref{eq:cap-state}) is the \textit{start} of a forward-time \textit{capture}.)

\subsubsection{Interplanetary transfer}
\label{methods:interplanetary-transfer}

The interplanetary transfer connects the Earth escape and Mars capture trajectories.
It starts at $X_{\mathrm{esc}}$ and ends along a~capture spiral that begins at some $X_{\mathrm{cap}}$.
Previous relevant work includes [e.g., \cite{hughes_realistic_2005},  \cite{tsuda_achievement_2013}, 
\cite{heiligers_optimal_2015},
\cite{song_solar-sail_2019}, \cite{sengupta_interplanetary_2026}].
We present a~solar-sail trajectory from a~specified low Earth orbit to a~specified low Mars orbit in which independently simulated Earth escape and Mars capture segments are joined by a~heliocentric transfer that matches the position and velocity (expressed in the heliocentric J2000 frame) and epoch at both interfaces.

For a~given escape state $X'_{\mathrm{esc}}$, we write the set of candidate interplanetary-transfer destination states from at $X'_{\rm esc}$ as

\begin{equation}
    \mathcal{S}_{X'_{\rm esc}} 
    \;=\;
    \bigl\{
    X_{\rm cap}\quad{\rm s.t.}\quad t_{\rm cap} \in [t_{\rm esc} + 500~{\rm d},\quad t_{\rm esc} + 1{,}300~{\rm d}]
    \bigr\}
    \label{eq:X_cap_set}
\end{equation}

For each $X_{\rm cap} \in \mathcal{S}_{X'_{\rm esc}}$ we seek a~trajectory that starts at $X'_{\rm esc}$ and ends in any of the states along the first 28 days of the capture spiral corresponding to $X_{\rm cap}$.
To meet this condition, the trajectory must deliver the sail to a~final state such that $|\Delta\mathbf r| < 10~{\rm km}$ and $|\Delta\mathbf v| < 1~{\rm m/s}$ relative to the desired final state at that epoch. Among the trajectories that meet this condition, we record the fastest.
If no such trajectory is found, the particular Earth escape state $X'_{\rm esc}$ is (given our assumptions) considered nonviable for interplanetary transfer. 
Appendix~\ref{app:interplanetary-traj-opt} has more details.

\subsection{Delivering sunlight to a human base}
\label{methods:delivering-sunlight}
We consider a human base on Mars close to 40$^\circ$N, 200$^\circ$E. This region is favored as a landing site given its low elevation, relatively low latitude, and shallow-subsurface ice [\cite{golombek_matthew_spacex_2021}].
For simplicity, we assume a Sun-synchronous, continuously sunlit, repeat-ground-track orbit.

\subsubsection{Constraints from orbit assumptions}
\label{methods:delivering-sunlight:constraints-from-orbit-assumptions}

We impose three constraints on our orbits:
(1) the~sun-synchronous condition constrains $i$ as~a~function of $a$.
(2) the~continuously sunlit condition constrains LTAN as a function of $a$. 
(3) The~repeat-ground-track condition constrains $a$.

The Sun-synchronous condition requires the~Right Ascension of the~Ascending Node (RAAN) to precess at an angular rate equal to Mars’ mean angular rate in orbit about the~Sun [\cite{brouwer_solution_1959}].
RAAN precession is driven by planetary oblateness, or $J_2$. We use $J_2 \approx 1.9566 \times 10^{-3}$, derived from Mars gravity field \texttt{MRO120F} [\cite{konopliv_detection_2020}].
The $J_2$-driven RAAN-precession rate varies with semimajor axis and inclination [\cite{brouwer_solution_1959}].
For the~altitudes considered ($\approx$ 500--1,450~km), the~permitted inclinations are all retrograde near-polar (93.21$^\circ$--96.77$^\circ$).

The Local Time of the~Ascending Node (LTAN), or the~local solar time at which the~sail crosses the~equator going northward, is constrained by our requirement that the~orbit be continuously sunlit (no eclipses).
The Sun-synchronous condition is defined using Mars' mean angular rate.
Because Mars' orbit about the~Sun is eccentric ($e \approx 0.09$), its instantaneous angular rate differs from its mean angular rate, so LTAN varies through the~year.
Over the course of a year, due to Mars' eccentricity, a Sun-synchronous orbit that begins with initial LTAN $L_0$ will drift between $\approx$ [$L_0-0.7$~h, $L_0+0.7$~h].
Because of this drift, we define the~interval $L^a_{\rm eclipse-free}$ as the~range of Mars-perihelion initial LTAN values for which Sun-synchronous orbits at~altitude $a$ (km) remain eclipse-free throughout the~year.
At 508 km altitude, $L^{508}_{\rm eclipse-free} \approx \{L_0 \in [17.38~{\rm h}, 18.38~{\rm h}]\}$\footnote{We find these intervals numerically; for more details, see Appendix~\ref{app:constellation-sizing}.}.
For simplicity, we use $L_0$~=~18~h and $a$~=~508~km as a representative orbit (see Table~\ref{tab:orbit-families}).

The repeat-ground-track condition requires an integer $k$ orbits per Mars solar day (88775.244 s).
This constraint simplifies the~calculations.
That is, in the~absence of perturbation by solar radiation pressure (SRP), we require that the~reflector would return to its original position and velocity in~the~Mars body-fixed frame (rotating, \texttt{IAU\_Mars}) after exactly $k$ orbits.
However, while the~reflector is reflecting light to the~human base, it cannot avoid experiencing solar radiation pressure perturbation.
The reflector must therefore use solar radiation pressure during the~non-illuminating parts of its orbit to counter this perturbation and maintain orbital stability.
Without active control, unless flying edge-on, low-Mars solar-sail orbits degrade quickly [\cite{bae_solar_2026}].

\subsubsection{Station-keeping in Mars orbit}
\label{methods:LMO:designing-for-orbital-stability}

We treat orbital stability as an optimization problem:
If the~sail starts at state $\mathbf x(t_0) = (\mathbf r_0, \mathbf v_0)$ with position $\mathbf r_0$ and velocity $\mathbf v_0$ at time $t_0$, we seek an attitude profile for the~sail such that the~difference between $\mathbf x(t_0)$ and $\mathbf x(t_1)$ is minimized, where ${\mathbf x}(t_1)$ is the~sail state one Mars day (sol) later.

We use ``attitude profile'' to refer to the~orientation of the~sail's normal vector $\nhat$ through time, denoted by $\nhat(t)$.
The sail orientation regulates solar radiation pressure acceleration: If $\shat$ is the~sail$\to$Sun unit vector, then $\nhat = \shat\implies$ maximum solar radiation pressure acceleration; $\nhat \perp \shat \implies$ zero solar radiation pressure acceleration.

To find an attitude profile that allows both orbital stability and human-base illumination, we first establish a baseline value for ``optimal illumination'' by propagating a point mass (unperturbed by solar radiation pressure) in the~reflector's candidate orbit, tracking the~visibility of the~human base from the~point mass.
When the~human base is visible to the~point mass (a ``delivery window''), we~find the~additional radiance (W/m$^2$) that would result at the~human base due to light from the~reflector (if it were in the same position as the point mass that is being propagated), assuming optimal pointing (Appendix~\ref{app:reflected-spot}).
(If optimal pointing would need too-fast slews or too much angular acceleration---Eq.~\ref{eq:esc-slew}---we discard that window).
We use the~resulting cumulative sol-averaged fluence (J/m$^2$), as the~baseline for optimal illumination.

Next, we optimize the~finite-agility attitude profile. 
Below, $u$ denotes the~argument of latitude, and $\nhat(u)$ the~attitude profile of the~sail.
Each iteration of optimization has four steps (Appendix~\ref{app:low-mars-orbit-optimization}):

\begin{enumerate}
    \item Propose a station-keeping-only attitude profile $\nhat^\dagger(u)$.
    \item Find a composite attitude profile $\nhat(u)$ including delivery windows, and smooth transitions (hereafter ``slews'') into and out of delivery windows.
    \item Propagate sail for one sol under $\nhat(u)$.
    \item Compute the~cost $\mathrm{J}(\nhat(u))$. $\mathrm{J}$ is greater for greater differences between initial and final states.
\end{enumerate}

\subsubsection{Long-term station keeping}
\label{methods:in-mars-orbit:multi-year-stability}

Solar radiation pressure varies $\approx1.5\times$ during Mars' year, given Mars' eccentricity---a destabilizing effect.
To ensure whole-year stability, we propagate the~sail for $\approx668$ sols (1 Mars year), re-optimizing the~attitude profile each sol. 
For some tests, we extend propagation to three Mars years.

We use each prior sol’s solution as a warm-start to the~subsequent sol’s optimization, which speeds computation and promotes (but does not enforce) smoothly changing solutions. 
The periodic formulation of the~attitude profile guarantees that, when optimizing $\nhat(t)$ for sol $i$, $\nhat(t_i) = \nhat(t_{i+1})$.
(Here $t_i$ is the~time at the~start of sol $i$, and $t_{i+1}$ is one Mars solar day later.)
But when constructing $\nhat(t)$ for the~next sol (sol $i+1$), the~optimizer may choose a solution with a different $\nhat(t_{i+1})$.
To~handle such cases, we impose a slew between the~end of sol $i$ and the~beginning of sol $i+1$ to respect the~attitude-control constraints of Eq.~\ref{eq:esc-slew}.

\subsection{Calculating energy delivery from a constellation of reflectors}
\label{methods:scaling-up}

An orbital plane, or ``ring,'' can deliver the~same energy\footnote{Ignoring mean-anomaly phasing discrepancies, which disappear in the~full-ring limit.} to a human base with a small number of bigger reflectors or a large number of small reflectors (Appendix~\ref{app:reflected-spot}).
Big individual reflectors simplify collision avoidance.
However, very big reflectors are ill-suited for our application.
Consider a single, giant reflector: it would only pass over the~base for $\approx15$~minutes, twice per sol. 
Smoothing energy delivery requires several dozen reflectors.
But even if such large reflectors ($\approx6~{\rm km}\times6~{\rm km}$) could be made, they could not be launched into orbit from Earth on a single launch vehicle. Moreover, bigger reflectors are harder to steer: their moment of inertia scales as the~fourth power of their side~length  [\cite{viale_attitude_2023}].

\citet{viale_reference_2023} proposed a reference architecture for OSRs: hexagonal spacecraft, 250~m per side (reflector surface area $\approx0.162$~km$^2$). 
As we want to show feasibility with near-term technology and support compatibility with a wide array of launch vehicles, we focus on smaller reflectors, which could conceivably be attitude-controlled with commercial, off-the-shelf (COTS)-like components.
Therefore, we consider $120~{\rm m}\times120{\rm m}$ square reflectors ($0.014$~km$^2$).
We do not present a spacecraft design for such a reflector, but we do consider constellation-sizing implications.
This choice of small reflectors motivates careful constellation design and introduces space traffic control challenges, which we discuss later (Section~\ref{disc:additional-challenges}).
Operationally collision-safe constellations around Mars may require larger individual reflectors to reduce the~total number of reflectors in orbit; this requires further investigation.

To deliver sunlight during more of the~sol, one must distribute reflectors across multiple orbital planes, or ``rings,'' at different LTAN values. 
A given altitude can only support so many rings before the~reflectors can no longer be phased in a collision-safe way; thus multiple altitude ``shells,'' each with many rings, must be used to deliver even more energy.
To size a realistically phased constellation we use a Walker-style approach [\cite{walker_satellite_1984}].
We consider altitude shells spaced every $\approx5$~km, ranging from the~$K_{12}$ reference altitude of 508~km up to 1432~km (100~km above the~$K_9$ reference altitude) and pack these shells with reflectors at different LTAN rings (Appendix~\ref{app:constellation-sizing}). 
We then estimate the~energy delivered by these reflectors as follows:

Let $f_{M_0,\,L_0,\,a}(t)$ denote the~instantaneous flux at the~human base in W/m$^2$ as a function of time $t$ from a single reflector with initial mean anomaly $M_0$, initial LTAN $L_0$, and altitude $a$ (ignoring atmospheric attenuation, discussed in Appendix~\ref{app:atmospheric-transmission}). 
We can then define the~fluence delivered to the~human base over one Mars solar day from that reflector as 
\begin{equation}
    F_{M_0,\,L_0,\,a}
    \;=\;
    \int_{t_0}^{t_1}f_{M_0,\,L_0,\,a}(t)\,\mathrm{d}t,
    \qquad \text{[J m$^{-2}$]}
    \label{eq:energy-from-one-reflector}
\end{equation}
where $t_0$ and $t_1$ are the~start and end epochs of that particular Mars solar day. Each reflector in an orbital ring has the~same $L_0$ and $a$, but has a different value of $M_0$. 
Thus we can find the~fluence from ring $R_{L_0,\,a}$ as:
\begin{equation}
    F^{\rm ring}_{L_0,\,a} 
    \;=\; 
    \sum_{M_0\in\mathcal{M}}F_{M_0,\,L_0,\,a},
    \qquad \text{[J m$^{-2}$]}
    \label{eq:energy-from-one-ring}
\end{equation}
where $\mathcal{M}$ is the~set of initial mean anomalies of the~reflectors in ring $R_{L_0,\,a}$. With $N_{L_0,\,a}$ reflectors in ring $R_{L_0,\,a}$, we define the~mean-anomaly-averaged fluence contribution from each reflector in ring $R_{L_0,\,a}$ as
\begin{equation}
    \bar{F}_{L_0,\,a} 
    \;=\; 
    \frac{F^{\rm ring}_{L_0,\,a}}{N_{L_0,\,a}}
    \qquad \text{[J m$^{-2}$ reflector$^{-1}$]}
    \label{eq:mean-anomaly-averaged-fluence}
\end{equation}
We compute $\bar{F}$ via simulation, but rather than simulate all $N$ reflectors for each ring, we sample $N=18$ reflectors equally spaced in mean anomaly for each ring:
\begin{equation}
    \bar{F}_{L_0,\,a} 
    \;\approx\; 
    \frac{1}{N}\sum_{k=0}^{N-1}F_{\frac{360k^\circ}{N},\,L_0,\,a}
    \qquad \text{[J m$^{-2}$ reflector$^{-1}$]}
    \label{eq:discrete-approx-mean-anomaly-avg-fluence}
\end{equation}
We then compute $F^{\rm ring}$ via scaling by the~actual number of reflectors in the~ring ($N_{L_0,\,a}$):
\begin{equation}
    F^{\rm ring}_{L_0,\,a} 
    \;\approx\;
    N_{L_0,\,a}\,\bar{F}_{L_0,\,a}
    \qquad \text{[J m$^{-2}$]}
    \label{eq:discrete-approx-ring-fluence}
\end{equation}
Assuming the~reflectors in a ring are equally spaced in mean anomaly, we~find that the~marginal fluence from a ring with $N+1$ reflectors compared to a ring with $N$ reflectors converges after $N\approx6$ (Fig.~\ref{fig:ring-to-ring-fluence}).
We thus conservatively sample with $N=18$. Each altitude shell has several rings, with each ring occupying a different LTAN. 
We compute the~fluence contribution from a shell $(F^{\rm shell})$ at~altitude $a$ by summing over its rings (represented by a set of ring LTAN values $\mathcal{L}$), and similarly for a constellation $(F^{\rm constell.})$ by summing over its shells (represented by a set of shell altitudes $\mathcal{A}$):
\begin{equation}
    F^{\rm shell}_{a} 
    \;=\; 
    \sum_{L_0\in\mathcal{L}} F^{\rm ring}_{L_0,\,a}\,
    \qquad
    \text{and}
    \qquad
    F^{\rm constell.}
    \;=\;
    \sum_{a\in\mathcal{A}} F^{\rm shell}_{a}.
    \label{eq:altitude-shell-fluence-and-constellation-fluence}
\end{equation}

\section{Results}
\label{sec:results}

\subsection{LEO to LMO feasibility}
\label{res:summary-LEO-to-LMO}

\begin{figure}[h]
    \centering
    \includegraphics[width=1\textwidth]{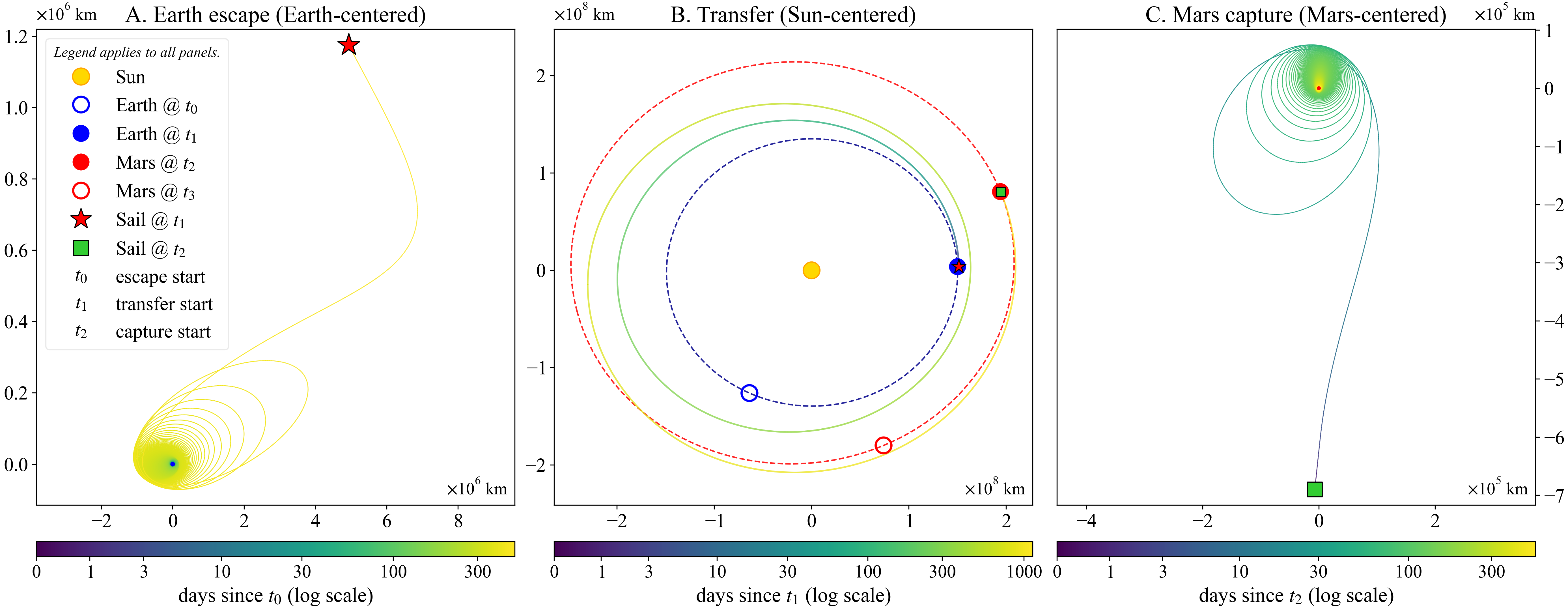}
    \caption{An example 3D interplanetary LEO$\to$LMO solar-sail trajectory, projected into 2D for visualization. 
    \textbf{A}: Earth escape, projected onto the escape's initial osculating orbit plane. 
    (Radial distances are in-plane; out-of-plane motion is foreshortened.) 
    \textbf{B}: Interplanetary transfer, projected into the J2000 X–Y (Earth mean-equatorial) plane. 
    \textbf{C}: Mars capture, projected onto the capture's best-fit plane. 
    Time $t_0$ denotes escape-start; $t_1$ denotes transfer-start; $t_2$ denotes capture-start. 
    Sail position at time $t_1$ denoted by \textcolor{red}{$\bigstar$} and at~time $t_2$ denoted by \textcolor{green}{$\blacksquare$}. 
    Color ramps show time since start of trajectory segment as a log scale.}
    \label{fig:transfer}
\end{figure}

To find the Earth escape duration $T_{\rm esc}$ as a function of areal density and departure epoch, we simulate escape (Section~\ref{methods:earth-escape}) each week for one Earth year using different orbiting solar reflector areal densities.
We find that $T_{\rm esc}$ grows nearly linearly with $\sigma$, at roughly 28~days per ${\rm g/m^2}$ (Appendix~\ref{app:additional-figures}, Fig.~\ref{fig:time_vs_sigma}).
This is because the solar radiation pressure acceleration scales as $1/\sigma$, so a heavier sail spirals out more slowly.
At $\sigma=18~{\rm g/m^2}$, $T_{\rm esc}\approx500$~days (range 479--557~days).
We do not consider time savings from, e.g., lunar gravity assist.

To find the Mars capture duration $T_{\rm cap}$, we model, for each $\sigma$ and for each of the four representative destination orbits (LTAN=18h, $M_0=0$ in each of $K_9$--$K_{12}$) (see Section~\ref{res:families-of-orbits-for-MMaaS}), biweekly arrivals over one Mars year.
$T_{\rm cap}$ grows nearly linearly with $\sigma$ but varies with capture phasing due to Mars' eccentricity and variable solar radiation pressure (Fig.~\ref{fig:time_vs_sigma}).
At $\sigma=18~{\rm g/m^2}$ the median capture takes $\approx 450$~days to $\approx 510$~days depending on destination-orbit altitude.
The lower the destination orbit, the longer capture takes.
Capture is feasible across the density range considered.

Patching escape, interplanetary transfer, and capture together at Earth's and Mars' Hill spheres produces a single continuous, flyable LEO~$\to$~LMO trajectory (Figure~\ref{fig:transfer}).
A $\sigma=18~{\rm g/m^2}$ solar-sail orbiting solar reflector departing from a polar, LTAN~=~18~h low Earth orbit reaches Earth's Hill~sphere after $\approx 487$~days; 
heliocentric transfer to Mars' Hill~sphere takes $\approx 1138$~days ($\approx 2.9$~years, varying significantly with Earth--Mars phasing); 
and capture delivers the reflector into a $K_{12}$ (508~km) low Mars orbit after another $\approx537$~days, for a total LEO~$\to$~LMO duration of $\approx5.9$~years.
The~transfer arc is the longest of the three legs, and its duration is set by the Earth--Mars phasing at departure [qualitatively consistent with \citet{hughes_realistic_2005}'s Fig.~3.2].
For $\sigma=18$~g/m$^2$, we find feasible end-to-end trajectories at launch dates every two months across one Earth--Mars synodic period.
This demonstrates that the LEO~$\to$~LMO delivery of a near-term buildable, slew-limited solar sail is possible, and~that delivery feasibility is relatively launch-window agnostic.

\subsection{Families of orbits for Mars-base support}
\label{res:families-of-orbits-for-MMaaS}
\begin{figure}[h]
    \centering
    \includegraphics[width=0.55\textwidth]{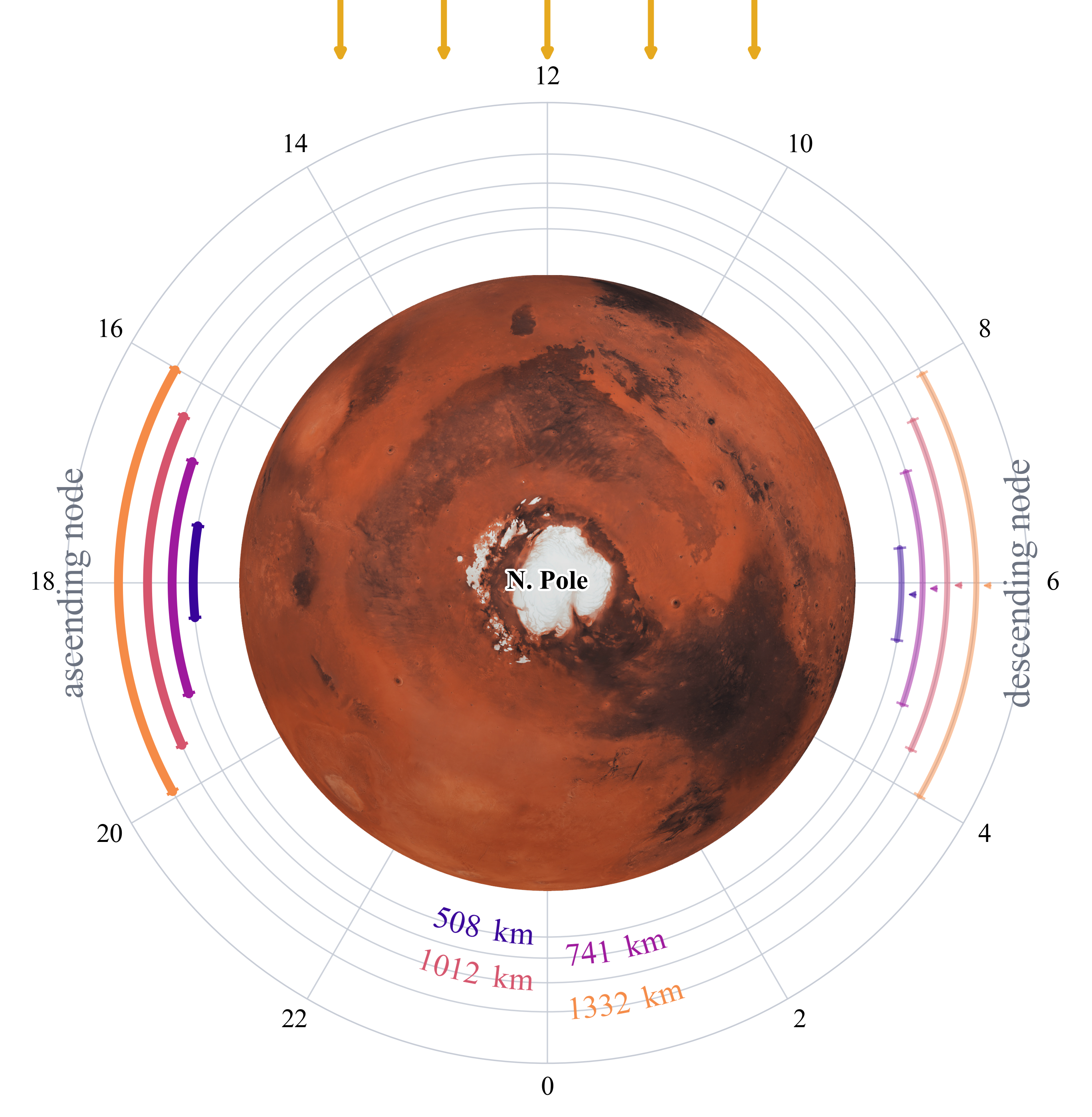}
    \caption{Candidate families of continuously sunlit Sun-synchronous orbits at Mars.
    OSR near-polar-orbit planes are out-of-page;
    sunlight enters from top of figure.
    Initial Local Time of the Ascending Node (LTAN) values and altitudes visualized with a~polar plot.
    Angular position denotes LTAN, while radial position denotes orbital altitude (to scale).
    The ranges $L^a_{\rm eclipse-free}$ for the repeat-ground-track altitudes $a \in \{508, 741, 1012, 1332\}$ km are shown on the left side of the figure.
    The corresponding local times of the descending nodes (LTDN) are shown on the right side of the figure. (Mars surface texture: solarsystemscope.com, CC BY 4.0)}
    \label{fig:feasible-LTANs}
\end{figure}

We find families of continuously sunlit, Sun-synchronous, near-polar, circular, retrograde, repeat-ground-track orbits (Table~\ref{tab:orbit-families}).
Each orbit family is numbered according to that family's particular repeat-ground-track altitude.
For instance, family $K_{12}$ corresponds to 12 revolutions per Mars solar day, and begins at $508$~km altitude.
The lower the orbit, the more tightly its initial LTAN ($L_0$) must be tuned to the dawn--dusk terminator to avoid entering Mars' shadow over the course of~a~year; a~higher orbit tolerates a~wider range of initial LTANs (nearly four hours).

\begin{table}[h]
\centering
\caption{\small
Four low-Mars orbit-families $K_9-K_{12}$.
Altitude is relative to Mars' equatorial radius $R_M = 3396.0$~km.
Inclination is the value needed for the Sun-synchronous condition at the given altitude, given all perturbations modeled for low Mars orbit (see Table~\ref{tab:forces-reference-frames}).
Eccentricity is set to zero for all orbits ($e=0$).
The eclipse-free LTAN range $L^a_{\rm eclipse\text{-}free}$ is the interval of initial local times of the ascending node for which the orbit remains continuously sunlit (no eclipse) over the full Mars year, including effects associated with Mars' declination and eccentricity.
Orbits are initialized at an epoch corresponding to Mars perihelion.}
\label{tab:orbit-families}
\begin{tabular}{lcccc}
\toprule
Family & Orbits per sol & Altitude (km) & Inclination ($^\circ$) & $L^a_{\rm eclipse\text{-}free}$ (h) \\
\midrule
$K_9$    &  9 & 1332.39 & 96.80 & 16.01--19.94 \\
$K_{10}$ & 10 & 1011.91 & 95.28 & 16.36--19.60 \\
$K_{11}$ & 11 &  740.81 & 94.20 & 16.74--19.15 \\
$K_{12}$ & 12 &  507.92 & 93.42 & 17.38--18.38 \\
\bottomrule
\end{tabular}
\end{table}

\subsection{Energy delivery and orbital stability metrics}
\label{res:delivered-energy-and-orbital-stability-metrics}
The orbital stability station-keeping algorithm (Section \ref{methods:delivering-sunlight}) produces long-term stable orbits with consistent delivery windows.
We initialize an OSR of areal density $\sigma=18~{\rm g/m^2}$ in the $K_{12}$ representative orbit (initial LTAN $L_0=18$~h, initial mean anomaly $M_0=0^\circ$, see Table~\ref{tab:orbit-families}) and propagate it 
for one Mars year (668 sols) under the optimized, station-kept trajectory that delivers sunlight to a single human base at $(40^\circ{\rm N}, 200^\circ{\rm E})$.
Fig.~\ref{fig:alpha-and-delta-one-sol} shows the attitude profile during the first sol of propagation.

\begin{figure}[ht]
    \centering
    \includegraphics[width=1\textwidth]{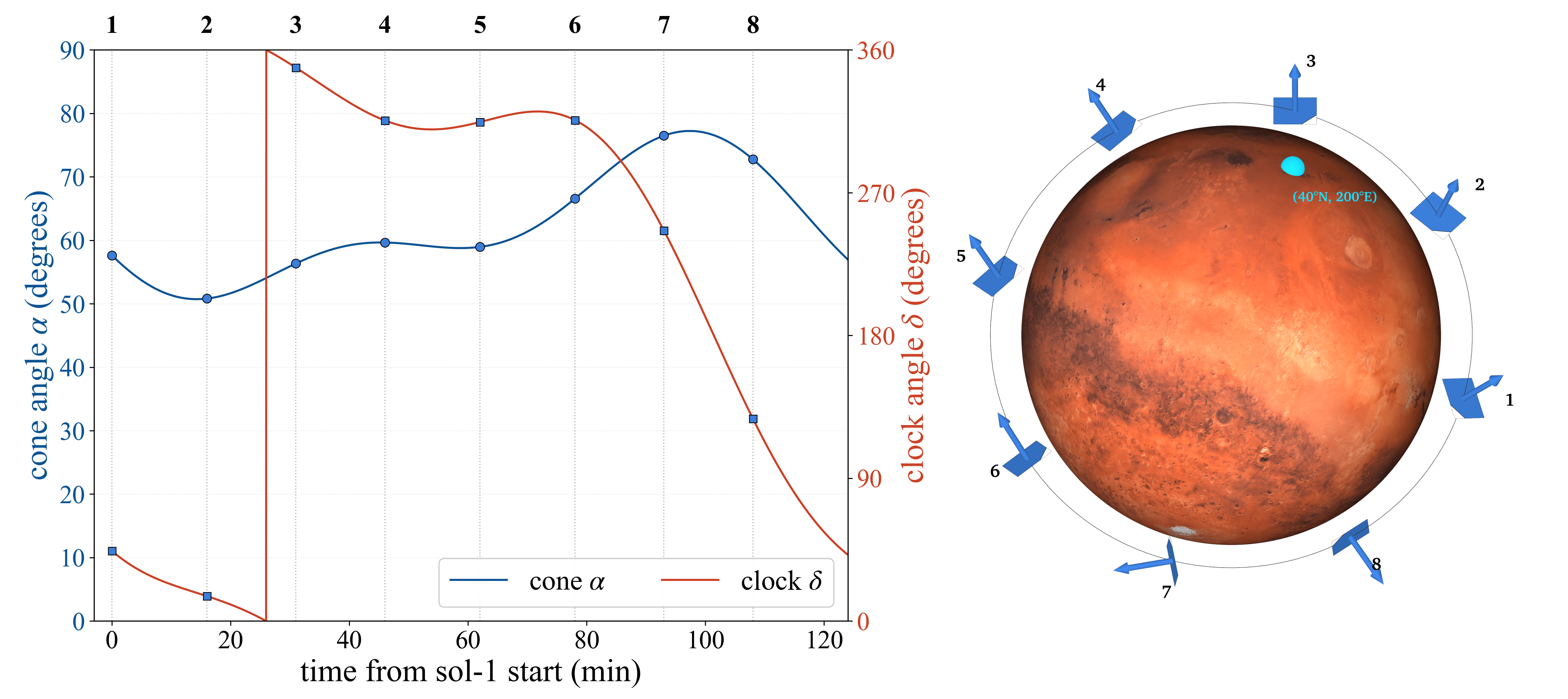}
    \caption{Representative orbiting solar reflector (OSR) attitude profile in 508-km altitude, circular, retrograde, near-polar orbit, during an orbit without a delivery window.
    In this orbit, the attitude profile is for station-keeping (i.e., orbital stability) only.
    Human base located at (40$^\circ$N, 200$^\circ$E) is marked with a~small blue dot.
    Orbiting solar reflector sketches are shown for eight points along the orbit.
    Arrows depict the normal vector $\nhat$ at each point.
    One corner of the reflector is highlighted to help visualize rotation.
    (Mars texture: \url{solarsystemscope.com}, CC BY 4.0.)}
    \label{fig:nonilluminating-orbit}
\end{figure}

\begin{figure}[h]
    \centering
    \includegraphics[width=1\textwidth]{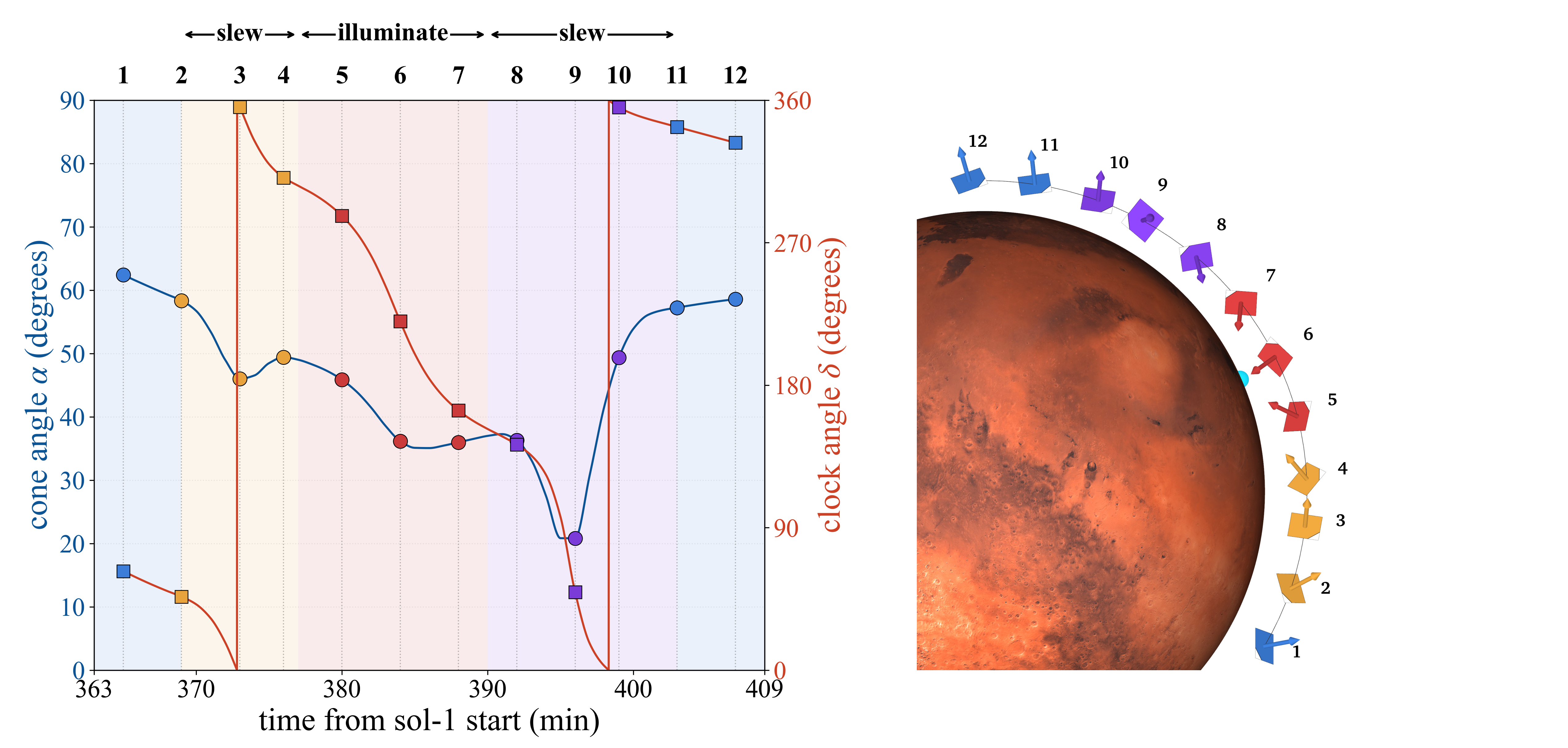}
    \caption{Orbiting solar reflector attitude profile during part of a 508-km altitude circular retrograde near-polar orbit that includes a delivery window.
    Human base located at (40$^\circ$N, 200$^\circ$E) is marked with a~small blue dot; local time: dusk.
    Orbiting solar reflector (OSR) sketches are shown for twelve positions. Colors correspond to stages of the orbit.
    \textbf{Blue:}~station-keeping.
    \textbf{Yellow:}~slewing into the illumination arc.
    \textbf{Red:}~optimal illumination of Mars-surface base ($\nhat$ bisects the Sun-sail-base angle).
    \textbf{Purple:} slewing out of the illumination arc to transition back to station-keeping. The vector pointing toward the Sun points out of the page toward the~reader. (Mars texture: \url{solarsystemscope.com}, CC BY 4.0.)}
    \label{fig:illuminating-orbit}
\end{figure}

\emph{Orbital stability}: \label{res:energy-and-stability:stability} Our implementation re-optimizes the attitude profile each sol to keep the~orbit closed. 
We measure closure, $\mathbf x(t_{i+1}) - \mathbf x(t_{i})$, in the Mars-fixed (\texttt{IAU\_MARS}) frame.
Each sol, the representative orbit closes tightly: the one-sol return residual (i.e., the displacement after one sol relative to the previous sol's handoff state) has a median of $0.84$~km and never exceeds $\approx 7$~km over the year, so the controller restores the repeat ground track on every sol.
The absolute body-fixed drift also stays bounded over the year: it grows to at most $\approx 142$~km near sol 200 before quickly reducing to $\approx 20$~km, oscillating between $0-20$~km until sol $\approx600$, when it begins to climb again.
Over the full Mars year, the~osculating eccentricity is always $<1.6\times10^{-2}$ (mean $5.7\times10^{-3}$).
That is, the reflector maintains a~near-circular orbit against the variable solar radiation pressure for an entire Mars year.

Figure~\ref{fig:nonilluminating-orbit} shows how sail orientation changes during an orbit without any delivery windows.
Attitude~reorientation here is solely for orbital stability, i.e., station-keeping.
Figure~\ref{fig:illuminating-orbit} shows how sail orientation changes during a delivery window, including the slews that transition out of and back into the station-keeping attitude profile.

\emph{Energy delivery}: \label{res:energy-and-stability:energy}
The orbiting solar reflector illuminates the human base twice per sol, every sol.
The per-sol fluence delivered to the base (from a representative $1{,}000~{\rm m^2}$ sail in a 508~km-altitude Sun-synchronous orbit with LTAN~=~18~h) averages to $25.5~{\rm J/m^2}$ ($\approx28~{\rm J/m^2}$ at perihelion, $\approx23~{\rm J/m^2}$ at aphelion).
The seasonal swing is modest even though the solar irradiance at the sail varies by $46\%$ ($1/r^2$ from $1.38$~AU to $1.67$~AU) (Appendix~\ref{app:reflected-spot}).
Mars' $25^\circ$ obliquity gives worse delivery geometry at aphelion for the $40^\circ$N human base.
Specifically, the angular separation between the Sun and the human base is greater, from the reflector's perspective, during northern-summer overflights.
Thus, for optimal pointing, the reflector must orient itself with a larger cone angle off the Sun-line, so less light is reflected to the base.

\subsection{Orbit altitude has modest effects on stability or energy delivery}
\label{res:impact-of-orbital-params-on-energy}

The optimization algorithm works for any repeat-ground-track orbit altitude.
Therefore, we use it to confirm long-term stability for $K_{11}$, $K_{10}$, and $K_9$, in addition to $K_{12}$.
Brighter sunlight is delivered from lower orbits: at perihelion, a single 1,000~m$^2$ OSR in an $L_0$~=~18~h $K_{12}$ orbit delivers $\approx 28~{\rm J/m^2}$ per sol to the human base, versus $\approx 16$, $\approx 15$, and $\approx 12~{\rm J/m^2}$ for $L_0$~=~18~h $K_{11}$, $K_{10}$, and $K_9$ orbits, respectively. 
This is due to the lower orbit's shorter slant range, which concentrates the reflected sunlight into a smaller, brighter spot.

\begin{figure}[h]
    \centering
    \includegraphics[width=0.8\textwidth]{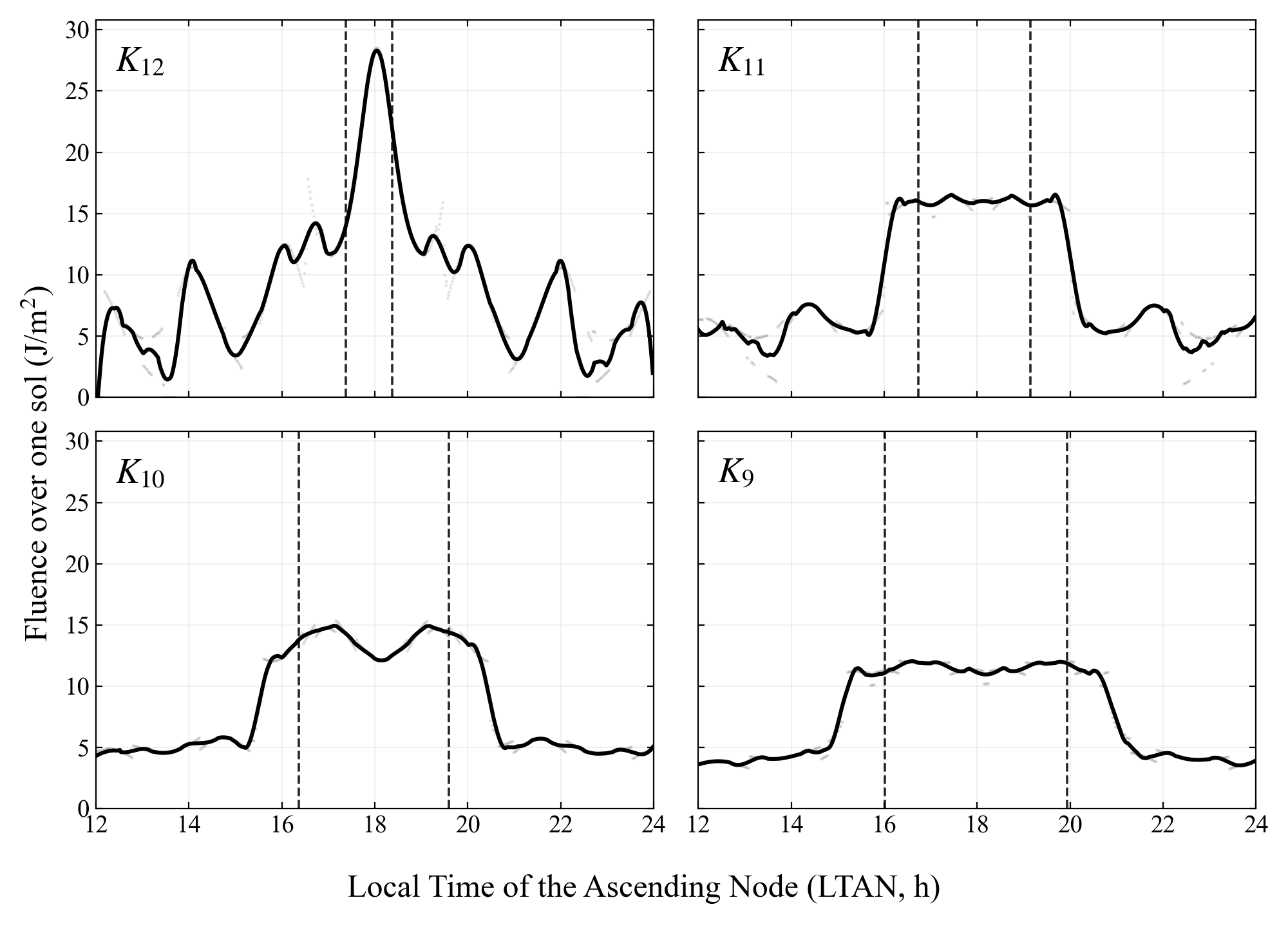}
    \caption{Fluence (J/m$^2$/sol) from a single $1{,}000~{\rm m^2}$ orbiting solar reflector with initial mean anomaly $M_0 = 0$, at Mars perihelion, as a function of initial LTAN ($L_0$). Panels correspond to the four repeat-ground-track altitudes considered (508~km, 741~km, 1012~km, 1332~km; see Section~\ref{res:families-of-orbits-for-MMaaS}). 
    Faint grey points plot the raw simulation output; black lines are fitted to these points and smoothed. 
    Dashed vertical lines indicate the~intervals of Mars-perihelion-epoch initial LTAN values that produce year-long eclipse-free orbits.}
    \label{fig:ltan-energy}
\end{figure}

\begin{figure}[h]
    \centering
    \includegraphics[width=1\textwidth]{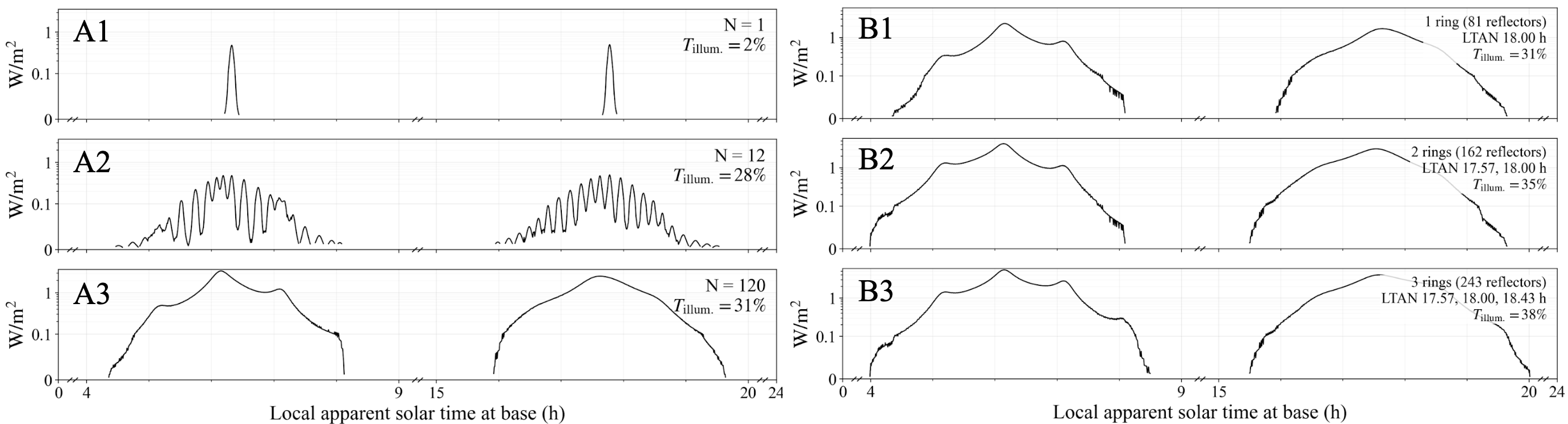}
    \caption{
    Irradiance delivered by OSRs to $(40^\circ{\rm N}, 200^\circ{\rm E})$ as a function of local solar time. 
    Reflectors are assumed to be square, $100{\rm m}\times100{\rm m}$, and simulation is done for Mars perihelion. 
    \textbf{A}-panels show the effect of adding reflectors to a ring (Section~\ref{methods:scaling-up} defines ``ring'' and ``shell''); the fraction of the day that has augmented sunlight ($T_{\rm illum.}$) increases with the number of reflectors in a ring ($N$) but plateaus at 31\% for this altitude. 
    \textbf{B}-panels show how adding rings to a shell can increase $T_{\rm illum.}$ beyond 31\%. 
    \textbf{A1}: irradiance from a~single reflector in the representative $K_{12}$ orbit ($L_0=18$~h, $M_0=0^\circ$). 
    \textbf{A2}: irradiance from a~ring of $N=12$ reflectors equally spaced in mean anomaly in the same $K_{12}$. 
    \textbf{A3}:~irradiance from a~ring of $N=120$ reflectors, also equally spaced. 
    \textbf{B1}:~irradiance from 81 equally spaced reflectors in an $L_0=18$~h, $a=508$~km ring. The fraction of the day during which reflectors illuminate the base is 31\%. 
    \textbf{B2}:~irradiance from two rings, each with 81 equally spaced reflectors: one~ring at $L_0=18$~h, and one at $L_0=17.57$~h. $T_{\rm illum.}=35\%$.
    \textbf{B3}:~irradiance from three rings, each with 81 equally spaced reflectors: one ring at $L_0=17.57$~h, one at $L_0=18$~h, one at $L_0=18.43$~h. $T_{\rm illum.}=38\%$.}
    \label{fig:densify-K12}
\end{figure}

Higher orbits offer robustness to LTAN, but lower peak intensity (Figure~\ref{fig:ltan-energy}).
Fluence from reflectors in $K_{12}$ (lowest orbit) peaks sharply near $L$~=~18~h within a~$\approx 1$~h eclipse-free band, whereas fluence from reflectors in $K_9$ (highest orbit) varies more smoothly within an eclipse-free band that spans nearly four hours.

The initial mean anomaly $M_0$ of a single reflector determines how many times per sol the reflector illuminates the base.
A $K_{12}$ reflector phased near $M_0\approx 0$ has two strong delivery windows, while one near $M_0\approx 180^\circ$ has four weaker windows.
However, for a full-ring constellation, this dependence washes out, as a densely and evenly phased ring of reflectors samples all $M_0$ values.
The relevant quantity, computed in Eq.~\ref{eq:energy-from-one-ring}, is the phase-averaged delivery.

\subsection{Scaling up: energy delivery from a constellation of reflectors}
\label{res:scaling-up-to-constellation}
An upper bound on the energy that a constellation of OSRs could deliver is the best single-reflector result---i.e., the peak per-sol fluence of a low-altitude orbiting solar reflector at its optimal ($L_0$, $M_0$) (Section~\ref{res:impact-of-orbital-params-on-energy})---multiplied by the number of reflectors in the constellation. This overstates the achievable energy:
$L_0$~$\neq$~18~h orbits deliver significantly lower fluence for the same initial mean anomaly $M_0=0^\circ$ (Fig.~\ref{fig:ltan-energy}), and realistic constellations would have reflectors at many altitudes, mean anomalies, and LTANs.

\begin{figure}[ht]
    \centering
    \includegraphics[width=0.9\textwidth]{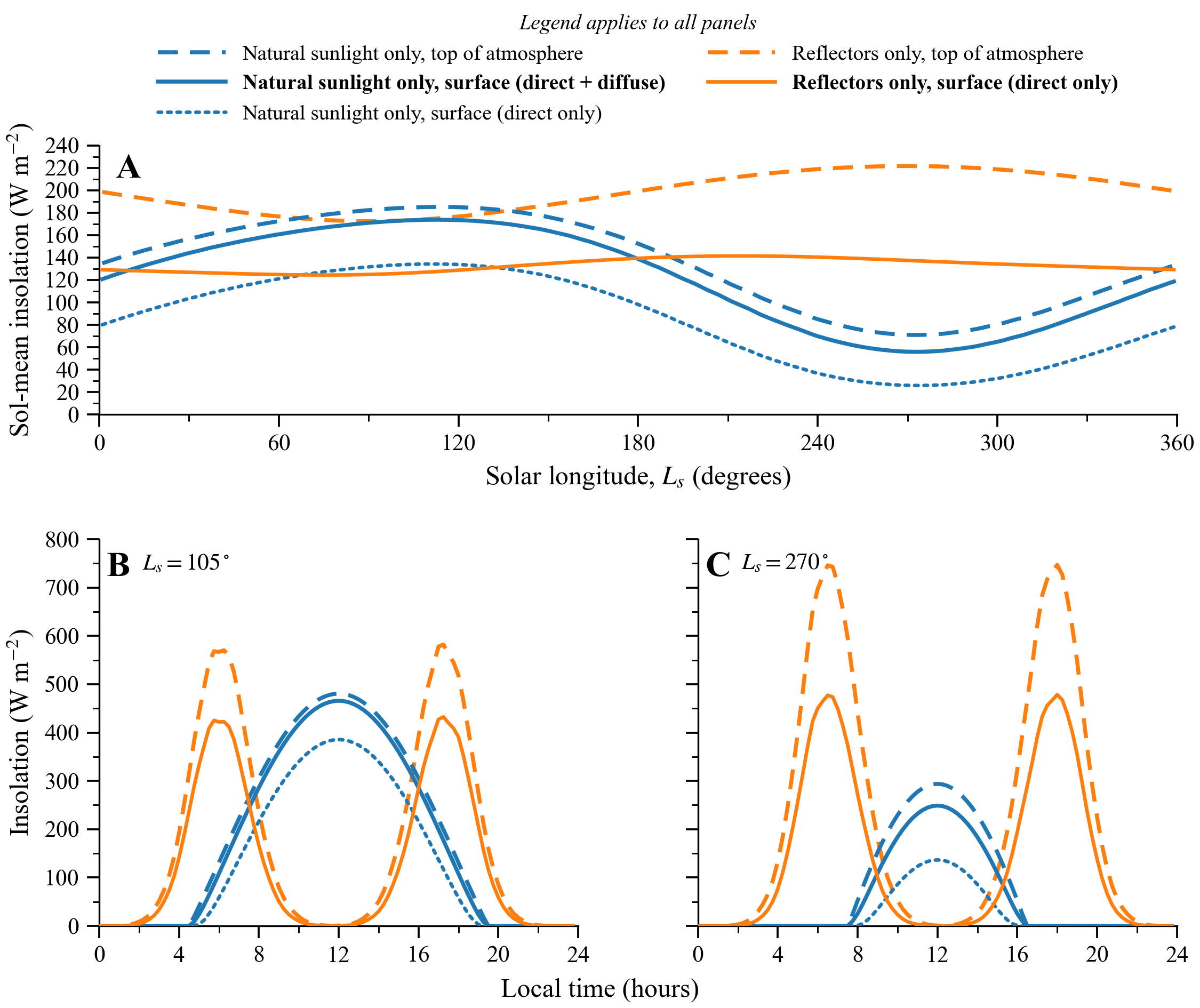}
    \caption{
    Insolation [${\rm W}\,{\rm m}^{-2}$] at a Mars base at ($40^\circ$N, $200^\circ$E) as a function of solar longitude $L_s$ ($^\circ$) and local time (h). 
    \textbf{A}:~Sol-mean insolation as a function of solar longitude. 
    \textbf{B}:~Instantaneous insolation during one sol at $L_s=105^\circ$ (minimum reflected-from-orbit contribution). 
    \textbf{C}:~Instantaneous insolation during one sol at $L_s=270^\circ$ (maximum reflected-from-orbit contribution).
    }
    \label{fig:multishell-power-to-base}
\end{figure}

Using the approach outlined in Section~\ref{methods:scaling-up}, particularly Eqs.~\ref{eq:energy-from-one-reflector}--\ref{eq:altitude-shell-fluence-and-constellation-fluence}, we find that a constellation with $\approx1{,}000$~km$^2$ reflector surface area in orbit could double the year-averaged insolation at a human base. 
One possible configuration for this constellation is $\approx10^5$ square OSRs, each $100{\rm m}\times100{\rm m}$ in~size, distributed across 185 altitude shells (Appendix~\ref{app:constellation-sizing}).
Ensuring operational collision safety may require that the constellation contain fewer, individually larger reflectors. 
For orbital rings with $N\gtrsim6$ reflectors, the phase-averaged fluence (Eq.~\ref{eq:mean-anomaly-averaged-fluence}) converges, and the power delivery from that ring becomes proportional to the total surface area of its reflectors (see Fig~\ref{fig:ring-to-ring-fluence}).
Table~\ref{tab:constellation-sizing-summary} gives the number of reflectors in each family of altitude shells for an example configuration. 
Figure~\ref{fig:multishell-power-to-base} shows the top-of-atmosphere irradiance (for both natural and reflected sunlight) at the human base.

\subsection{Microclimate modeling results}
\label{res:microclimate-modeling-results}

\begin{figure}[h]
    \centering
    \includegraphics[width=0.9\textwidth]{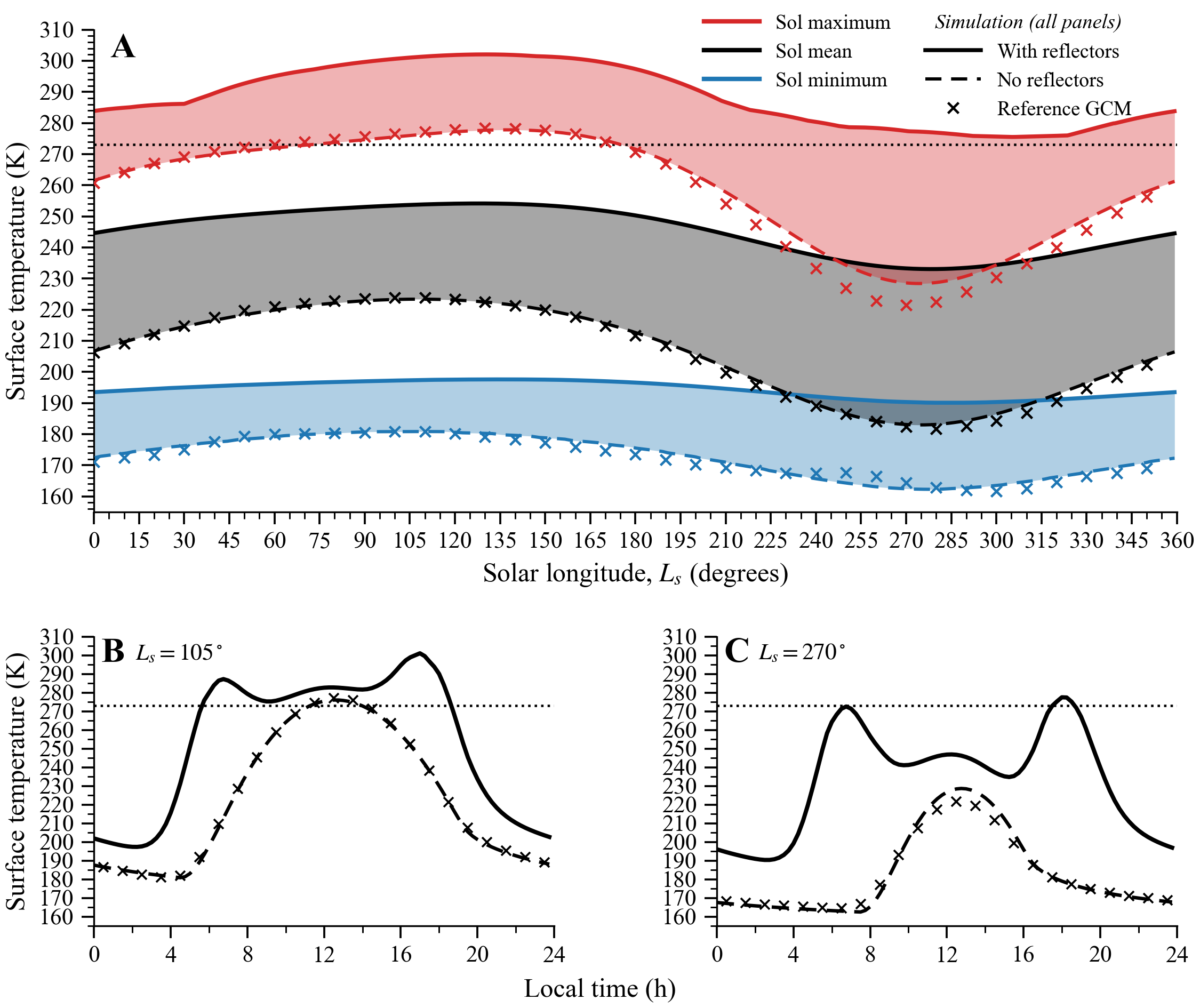}
    \caption{Temperature with (solid) and without (dashed) additional insolation from orbiting solar reflectors. Horizontal dotted line shows the melting point of water ($273~{\rm K}$). \textbf{A}:~\textcolor{blue}{Minimum} (blue), mean (black), and \textcolor{red}{maximum} (red) temperatures each sol at (40$^\circ$N,~200$^\circ$E) as a function of solar longitude ($L_s$). 
    Reference values from the Global Climate Model (Appendix~\ref{app:1D-climate-model}) plotted as crosses. 
    \textbf{B}:~Temperature during one Mars day at $L_s=105^\circ$. \textbf{C}:~Temperature during one Mars day at $L_s=270^\circ$.
    }
    \label{fig:temp-vs-Ls}
\end{figure}

We estimate the surface temperature at the human base with and without a constellation of OSRs using a one-dimensional climate model by \citet{kling_subsistence_2020} to solve the surface energy balance:

\begin{equation}
\frac{\partial U}{\partial t}
\;=\; 
(1-a) I_{\rm total} + 
Q_{\rm IR \downarrow} + 
Q_{\rm cond} +
Q_{\rm geo} - 
\mathit{SH} - 
\epsilon \sigma T^4_{\rm surf} 
\label{eq:climate-model}
\end{equation}

where $U$ is the energy per unit of area for the surface layer, $a$ the surface albedo, $I_{\rm total}$ the solar irradiance at the surface from the reflectors and the Sun including attenuation by the atmosphere (Appendix~\ref{app:atmospheric-transmission}), $Q_{IR \downarrow}$ the downward longwave radiation from the atmosphere, $Q_{\rm cond}$ the conductive flux through the surface, $Q_{\rm geo}$ the geothermal flux, $\epsilon \sigma T^4_{\rm surf}$ the longwave upward radiation with emissivity $\epsilon \sim$ 1, $\sigma$ the Stefan--Boltzmann constant, and $\mathit{SH}$ the sensible heat lost to the atmosphere. 

Appendix~\ref{app:1D-climate-model} details the terms in Eq. (\ref{eq:climate-model}). 
We validate the one-dimensional (1D) climate model against a 3D Global Climate Model (GCM) simulation by comparing the daily mean, maximum and minimum temperatures at 40$^\circ$N from the GCM (zonally-averaged at 40$^\circ$ to remove transient atmospheric features), and from the one-dimensional climate model with no reflectors (respectively the $\times$-markers and the dashed line in Fig.~\ref{fig:temp-vs-Ls}). 
The daily-averaged, annual-mean surface temperature from the one-dimensional climate model are $\approx208~{\rm K}$ for both (within $0.5~{\rm K}$ agreement). 
The 1D~climate model also reproduces well the daily minimum and daily maximum (Fig.~\ref{fig:temp-vs-Ls}A) temperatures from the GCM. 
Disagreement is worst $\approx7~{\rm K}$ in daily maximum temperature during the high-dust season ($L_s \approx 270^\circ$, Fig.~\ref{fig:temp-vs-Ls}C) when the one-layer parameterization for atmospheric scattering has the largest divergence with the GCM's (which leverages a computationally extensive multi-layer radiative transfer model). 
At other seasons the 1D~climate model matches the diurnal temperature cycle from the GCM (e.g Fig.~\ref{fig:temp-vs-Ls}B), and therefore provides a suitable baseline to study the effects of the OSR on the surface temperatures.  

Without reflectors, the mean temperature is $208~{\rm K}$ ($\min=183~{\rm K}$ at $L_s=280^\circ$, $\max=223~{\rm K}$ at $L_s\approx 105^\circ$, and daily maximum temperature only exceeds the freezing point of water ($273~{\rm K}$) on 223 out of 668 sols (1/3) in the Mars year.
With the additional insolation delivered to the base by the OSR constellation, the mean local temperature rises to $246~{\rm K}$ ($\min=233~{\rm K}$ at $L_s=270^\circ$, $\max=254~{\rm K}$ at $L_s=130^\circ$), a $+38~{\rm K}$ increase in the annual mean value. The daily-maximum temperature exceeds the freezing point of water every day for the entire year and the absolute minimum temperature never drops below 190~K (vs 162~K without reflectors).

\section{Discussion}
\label{sec:discussion}

\subsection{Pointing accuracy requirements are modest}
\label{disc:sensitivity-to-pointing-error}
Our calculations of energy delivery assume optimal pointing.
In reality, a reflector's pointing accuracy is imperfect [\cite{moore_effects_2024}].
How sensitive is energy delivery to pointing error?
From the~specular-reflection geometry (Eq.~\ref{eq:app-bisector}), for a given deviation $d\theta$ from bisector pointing, the~reflected spot's center shift, $dr$, is proportional to the~slant range $d$ (and thus grows with altitude). A normal tilt $d\theta$ deflects the~reflected ray by $2\,d\theta$, so the~spot center moves $dr \approx 2\,d\theta\,d$ (with $d$ approximately the~altitude when the~sail is closest to being overhead from the~base's perspective).
There is least margin for pointing error when the~reflected spot size is smallest. 
The spot is smaller for lower orbits (Eq.~\ref{eq:app-spotarea}); when the~reflector is closest to being overhead; and when Mars is farthest from the~Sun (because the~spot shrinks with the~Sun's angular diameter).

Our worst-case analysis is therefore for the~lowest orbit we consider ($\approx$ $500$~km altitude), at aphelion, and at the~resulting peak elevation $\varepsilon \approx 66^\circ$ (recovered from simulations). A pointing deviation of $d\theta = 1$~mrad from optimal pointing moves the~reflected spot by $\approx1.1$~km.
In this smallest-spot case (for our representative orbit), the~reflected spot has semimajor axis $\approx1.7$~km and~semiminor axis $\approx1.6$~km (area $\approx8.3$~km$^2$).
For a $1$~km$^2$ circular solar array at the~human base, the~margin for pointing error is $\approx0.9$~mrad before the~solar array is~no~longer~contained within the~reflected spot.
This worst-case is still a modest requirement compared to flight-demonstrated reaction-wheel-based pointing accuracy [e.g., \cite{karpenko_first_2012}].

\subsection{Tension requirement increases spacecraft mass}
\label{disc:tension-and-buildability}

The reflected spot size ($\gtrsim\!8$~km$^2$) from an OSR is bigger than the predicted solar farm for an initial Mars base.
Thus, some reflected light is wasted in the context of solar-power delivery.
(For~microclimate modification or melting shallow-subsurface ice, the efficiency may be greater.)
Sail~wrinkles waste even more sunlight.
For example, a surface slope error of~$\approx\!1$~mrad ($\approx\!0.06^\circ$) would deflect the reflected rays by $\approx\!2$~mrad, broaden the beam's angular radius to $\approx\!5$~mrad, and triple the spot size, from $\approx\!8$~km$^2$ to $\approx\!25$~km$^2$ (Eq.~\ref{eq:app-spotarea}). To keep the~reflector membrane flat, tension is needed to smooth folds that persist after deployment, prevent the membrane from deforming as it reorients, and counter solar radiation pressure billow.
With a~square deployable solar sail,
high tension in the membrane requires larger booms to prevent buckling.
This in turn increases the reflector's moment of inertia, requires a more capable attitude control system, and~overall increases non-sail mass. Thus, the tension requirement increases the~sail areal density.

\subsection{Cost and benefits}
\label{disc:oom-cost-estimates}
To get a rough estimate of costs and benefits, we calculate the total surface area of OSRs needed to double insolation at a Mars base and find the corresponding constellation mass. 
Multiplying by the estimated cost per kilogram to~make and launch to 800~km Earth polar orbit (both of which are very uncertain, because we do not know how far launch prices will fall) gives constellation cost. 
From this Earth orbit, we assume the reflectors will then solar-sail to Mars.
We define $c_{\rm launch}$ as the launch cost per unit mass [\$/kg] to 800~km Earth polar orbit, $c_{\rm make}$ as the manufacturing cost per unit mass [\$/kg], and $\sigma$ as the constellation mass divided by the constellation reflector surface area [g/m$^2$]. The cost $C$ of the constellation (considering only procurement and launch) is then

\begin{equation}
    C \;\approx\; A \,\sigma^\star\,(c_{\rm launch} + c_{\rm make})
    \label{eq:cost-estimate}
\end{equation}

where $A$ is the total reflector area in m$^2$ and $\sigma^\star = 10^{-3} \times \sigma$ (to convert from g$\to$kg). 
The constellation has total area $A \approx1.4\times10^9~{\rm m}^2$. 
We assume future cost-to-orbit $c_{\rm launch}=\$250$/kg for the 800~km low Earth orbit. 
For procurement cost, \citet{viale_reference_2023} estimate $\approx\$350$/kg for OSRs that are individually larger ($\approx10^5$~m$^2$ each, as opposed to $\approx1.4\times10^4~{\rm m}^2$ each), assembled-in-orbit OSRs. 
These are similarly light (18 g/m$^2$), and capable of similar slews as those we consider in this work. 
We set $c_{\rm make} = \$350$/kg. These assumptions give $C\approx\$42{\rm B}$ at $\sigma = 50~\rm{g/m^2}$ and $C\approx\$13{\rm B}$ at $\sigma = 15~\rm{g/m^2}$, although the true cost depends on the future trajectory of cost-to-orbit, which is unknown.

A constellation of OSRs can illuminate many bases.
Even when tasked to illuminate three bases equally spaced in longitude, all at $40^\circ$N, with a greater number of delivery windows and corresponding slews, we find that the remaining time is enough for the OSR to maintain long-term stability.
Letting the cost per base when $b$ bases get served by a single constellation be denoted by $C_b$, at $\sigma=15{\rm g/m^2}$ and $b=3$, $C_b \approx \$4{\rm B}$.

\textit{Synergies with Mars solar farms and Mars satellite delivery}: Suppose a future where there is demand for 10--100~MW power from one or more bases on Mars. 
To meet the upper range of this power demand with only solar panels and batteries would require $\approx10$~km$^2$ in installed panel area\footnote{We assume 16\% panel efficiency, consistent with \url{https://www.starlight.space/starlight-air-datasheet.pdf}; base(s) at 40$^\circ$N; and panels placed flat on the ground.}. 
We do not know how much such a power system would cost if it were made on Earth and shipped to Mars. 
Using estimates that assume a 40-fold decrease in cost to Mars surface relative to current prices,\footnote{We assume launch costs of \$2,500/kg to Mars' surface; balance-of-plant (BoP) mass equivalent to $2\times$ panel mass; BoP procurement cost of \$50/kg; panel areal density of 0.8~kg/m$^2$, and panel procurement cost of \$3,250/m$^2$, consistent with \url{https://www.starlight.space/starlight-air-datasheet.pdf}.} the dollar-per-Watt cost would be $\approx\$410$/W, not including batteries.

However, with a constellation of Mars-orbiting solar reflectors enhancing solar power generation, this power demand could be met with less than half the installed panel area.
The combined system ($\approx1.4\times10^3~{\rm km}^2$ of reflector surface area in orbit and 3.2~km$^2$ of solar panels on the surface) would cost\footnote{We assume reflector areal density of 18~g/m$^2$; reflector procurement cost of \$350/kg (consistent with \citealt{viale_reference_2023}); and to-LEO launch costs of \$250/kg.} $\approx\$260$/W, again not including batteries.
When we consider energy storage, relative costs for the with-OSRs scenario at high power demand decrease further.
This is because the requirements for energy storage would be approximately halved if OSRs are used, as the reflectors would enable power generation at night (Fig.~\ref{fig:multishell-power-to-base}).

Power systems for a solar-powered Mars base occupied year-round will be designed for the worst-case illumination, which at 40$^\circ$N is near $L_s$~=~270$^\circ$ (Fig \ref{fig:multishell-power-to-base}).
At $L_s$~=~270$^\circ$, the sol-averaged surface flux (after attenuation by dust) is 56~$\mathrm{W~m^{-2}}$ from natural sunlight and 138~$\mathrm{W~m^{-2}}$ from the reflectors alone, totaling 194~$\mathrm{W~m^{-2}}$ together. 
Those fluxes can be rewritten in units of $\mathrm{MW~km^{-2}}$ (1~$\mathrm{W~m^{-2}}$ = 1~$\mathrm{MW~km^{-2}}$) to calculate the daily energy production potential at the site: 
$E^{\rm nat}_{\rm tot}=56~{\rm MW~km^{-2}}\times24~\mathrm{hr}\approx1.3\times10^3 \;  \mathrm{MWh~km^{-2}}$ with only natural sunlight and 
$E^{\rm nat\,+\,OSR}_{\rm tot}= 194~{\rm MW~km^{-2}} \times 24~\mathrm{hr}\approx 4.7\times10^3 \; \mathrm{MWh~km^{-2}}$ with the addition of OSRs (a~factor of $\approx3.6$), excluding solar panel efficiencies.
(For clarity, we round the Mars-sol duration to 24~hr. It is really $\approx 24.66$ Earth hours; the difference only changes the estimates by a few percent.)

An important consideration in sizing the power system is the size of the storage battery which must supply power after sunset. 
In turn, this depends on the activity planned at the base as a function of the time of day. 
For simplicity, we assume a constant power consumption at the base $P_o$. 

While the sizing of the solar panel area $A$ for a given power consumption $P_o \; [{\rm MW}]$ depends only on total energy production via $A=P_o/\mu I_o \;\mathrm{[km^2]}$ (where $\mu=16\%$ is the assumed solar panel efficiency\footnote{consistent with \url{https://www.starlight.space/starlight-air-datasheet.pdf}} and $I_o$ the sol-averaged insolation), the sizing of the corresponding battery system depends on the insolation profile as a function of the time of day. 

A useful metric is the earliest local time of day when 100\% of the power is supplied by the solar~power~system:
\begin{equation}
   LT_{100\%} ~=~\text{first local time of day such that}~~I(t) \mu A > P_o \;\; \mathrm{[MW]}.
\end{equation}
After replacing $A=P_o/\mu I_o$ in the equation above, we observe that this local time is solely dependent on the insolation profile and occurs when $ I(t)/I_o > 1$.  
At $L_s$ 270$^\circ$, $LT_{100\%}^{\rm nat}$ is 08:30 with only natural sunlight and $LT_{100\%}^{\rm nat\,+\,OSR}$ is 04:45 with the addition of the OSRs (independent of power consumption). 
Thus with OSRs, the solar array can fully power the base for a longer part of the day, so a smaller battery is needed. 

As sunlight delivered (natural and reflected) is symmetric around noon (Fig.~\ref{fig:multishell-power-to-base}), the usable battery capacity required to deliver constant power $P_o$ can be written as  
\begin{equation}
   C_u = 2 \eta   P_o\cdot LT_{100\%}~~ \mathrm{[MWh]}
   \label{eq:battery}
\end{equation}
The factor of 2 in Eq.~\ref{eq:battery} implies that the battery storage system must provide power for similar periods of time before sunrise and after sunset (in other words, usable battery capacity is  $\approx$50\% in the middle of the night). 
$\eta$ is a tuning factor close to unity and based on the exact power delivery profile for natural sunlight and with OSR, used to ensure that the battery charge never falls below 0\% at any time, and that the charge level at the end of the cycle is no less than at the beginning of the cycle. 
We find $\eta^{\rm nat}=0.96$ and $\eta^{\rm nat+OSR}=0.88$ provide a charge level $\ge$ 50 \% at the end of the cycle. We assume a 20\% margin on battery capacity so that charge levels never drop below 20\%.
The value in $C_u$ describes usable capacity. 
Table \ref{tab:table_cost} provides estimates for the size and costs of the system for various target power generation levels for the base.

\begin{table*}
  \centering
  \small
  \begin{tabular}{cccc|cccc}
    \toprule
    \multirow{2}{*}{Production [MW]}& \multicolumn{3}{c|}{\textbf{Natural Sunlight}} & \multicolumn{3}{c}{\textbf{Natural Sunlight + OSRs}} \\
    \cmidrule{2-7} & $A$ [km$^2$] & $C_u$ [MWh] & Cost [\$B] & A [km$^2$] & $C_u$ [MWh] & Cost [\$B]  \\
    \midrule
    10 & 1.1 & 163 & 13 & 0.3 & 84  &  19 \\ 
    \midrule 
    50 & 5.6 & 816 & 69 & 1.6 & 418 &  38 \\ 
    \midrule 
    100& 11.1& 1632& 138& 3.2 & 836 &  62 \\ 
   \bottomrule
  \end{tabular}
  \caption{Solar array surface area ($A$), usable battery capacity needed to serve power demand at times-of-day when solar is not available ($C_u$), and cost estimates for the full system, for three different power-demand scenarios (10~MW, 50~MW, and 100~MW). 
  In each power-demand scenario, the full constellation with $\approx1{,}400~{\rm km}^{2}$ reflector area is assumed to be present, with its costs included in the cost estimate.
  We assume 80\% of the battery's nameplate capacity is usable, i.e., $C_{\rm nameplate}=C_u/0.8$, leaving a 20\% minimum state-of-charge reserve. We assume 16\% solar panel efficiency, 150~Wh/kg battery energy density, \$164/kWh battery procurement costs, and Mars shipping costs for batteries equivalent to those for solar panels (Eq.~\ref{eq:cost-estimate}).  
  Analysis is based on solar longitude $L_s$~270$^\circ$.}
  \label{tab:table_cost}
\end{table*}

Effective OSR system cost could also be reduced by transporting freight to Mars.
For example, if~a~$120~{\rm m}\times120~{\rm m}$ reflector is nominally 18~g/m$^2$, each reflector could support up to 170~kg in freight and remain below 30~g/m$^2$ in combined areal density, or $460$~kg in freight while remaining below 50~g/m$^2$. 
Freight could be centrally mounted to~reflectors and thus contribute negligibly to the reflector's moment of inertia. 
This way, the reflector could still perform the maneuvers needed for LEO$\to$LMO flight, but at the cost of longer transit time.

\subsection{Additional challenges and model limitations}
\label{disc:additional-challenges}
Flight tests are needed before a constellation of Mars-base-warming OSRs can be realized. 
One example [\cite{kite_research_2026}] would be a LEO$\to$LMO pathfinder with a low-areal-density solar sail [\cite{hughes_realistic_2005}]. 
For such a pathfinder, the~tension requirement for the~sail membrane could be relaxed, making it easier to~achieve low areal density.

Experiments on the~relationship between tension and surface error [e.g., \cite{blandino_corner_2002}, \cite{wong_wrinkled_2006}, \cite{bonin_-wrinkling_2014}, \cite{zou_experimental_2022}] in~the very-low-surface-error regime of a precise OSR would be valuable.

A spacecraft design that satisfies the~requirements for a Mars-base-warming OSR---low areal density, high tension in the~sail membrane, agile maneuverability, longevity, and affordability---is needed. Such a design might repurpose consumer electronics components (e.g.,~cellphone-camera sensors for star-tracking) [\cite{handmer_how_2024}]. 
However, more robust components may be needed to address the radiation dose accumulated while spiraling up through Earth's Van~Allen belts.
Other challenges include:

\textit{Pointing and navigation}: We assumed that each reflector has perfect knowledge of~its~state.
In reality, imperfection arises from (e.g.) the~precision of each reflector's star tracker, optical sensors, and inertial measurement unit; the~fidelity and sensitivity of the~actuators in each reflector's attitude control system; and deviations from the~flat-plate ideal model of each reflector due to~flex or vibration.
Near Earth, spacecraft can use Earth's magnetic field to~help fix position and orientation; this does not work at Mars, as Mars lacks a global magnetic field.
For actual flight missions, the~robustness of the~trajectories and orbital stability management strategies proposed here should be assessed, and uncertainty-aware trajectories and orbits designed.

\textit{Space traffic control and communications}: If a higher-bandwidth Earth--Mars communication infrastructure is not available, the~reflectors would need to~be mostly autonomous; or communicate with Mars-based ground stations; or be managed by compute-and-communication hub satellites.

\textit{Self shadowing}: Reflectors at lower orbits might pass through the beams of reflected light from reflectors at higher orbits, partially blocking those beams for short durations.
This effect is likely small and we did not model it, but further investigation is needed to quantify this effect.

\textit{Managing failed spacecraft}: Random events could render individual reflectors uncontrollable.
A~debris strike that fragments a~reflector could be catastrophic for part of the~constellation, depending on how the~fragments' orbits evolve. 

Loss of control (with no fragmentation) could cause the~sail to~spin rapidly, spin slowly, or freeze in a~fixed orientation. 
We considered each case and simulated (for $\sigma~=~18~{\rm g/m}^2$) how the~reflector's orbit would evolve.
At the~$K_{12}$ altitude (508~km), fixed orientation causes the~reflector to~deorbit (intersect Mars' atmosphere) after just 16~days.
Slow tumbling (1$\times$/orbit) causes the~orbit to~intersect the~atmosphere within months.
Fast tumbling (1$\times$/minute) does not result in intersecting the~atmosphere within two Mars years.
For $K_{11}$, $K_{10}$, and $K_{9}$, slow tumbling did not lead to deorbiting, and fixed pointing took 35, 58, and 270 days, respectively, to~reach the~top of~the~atmosphere. 

For fragmented spacecraft, solar radiation pressure would likely swiftly deorbit sail membrane fragments (via eccentricity pumping, causing intersection with the~atmosphere), due to their low areal density. Higher areal density fragments such as the~spacecraft bus, reaction wheels, or the~carbon-fiber booms pose greater risk to~the constellation. 
Redundant attitude-control or localized debris shielding on high-risk areas could mitigate risk, but at the expense of higher~mass, and may prove infeasible in practice.

More modeling of the~orbits of fragments of different sizes and densities would help bound the~risk [\cite{suchantke_space_2020}]. 
Constellation safety requires knowing the~critical number of intact objects that triggers a collisional runaway.
Such~orbital safety modeling is performed for LEO constellations [\cite{lewis_critical_2026,kessler_critical_2001}]. 
A similar model developed for~Mars is needed to~understand how many sails could be safely deployed.
Further developments in active debris removal would also improve constellation safety [e.g., \cite{aglietti_active_2020}, \cite{godfrey_beyond_2024}].
The sustainability of low Mars orbit [cf. \cite{suchantke_space_2020}'s Mars Sustainability Framework] must also be considered for constellation design.

\emph{Other limitations}: Our simulation omits atmospheric drag at Mars (except for a~300~km altitude ``floor''), and omits OSR membrane degradation (e.g., from solar wind). 
Moreover, the~eccentricity growth that we report in Section~\ref{res:energy-and-stability:stability}, while bounded and small, is likely still too great for such a dense constellation that relies on circular orbits. 
These details matter when designing an OSR constellation for multi-decade operations. 
We do not consider in-situ resource utilization production of solar panels at Mars, nor of orbiting-reflector components at Mars' moons. 

\subsection{Alternative Mars orbits}
\label{disc:alternative-mars-orbits}

We consider only circular, near-polar, Sun-synchronous, continuously sunlit orbits. 
Alternative orbits could also warm a Mars base.

To keep reflectors in sunlight, we choose initial LTAN ($L_0$) so that the orbit avoids eclipses year-round.
The natural variation in LTAN due to Mars' eccentricity then permits only a small range (1~h) of $L_0$ for the lowest-orbiting reflectors.
A wider $L_0$ range is possible if reflectors modify their LTAN using solar radiation pressure, or have batteries for eclipse robustness.
Either approach would allow larger constellations, and reflected sunlight at the bases later after sunset (and earlier before dawn).

Similarly, if solar radiation pressure is used to help drive RAAN precession, that would partly relax the constraint on inclination imposed by Sun-synchronicity and unlock more orbital inclinations. 
Less clear is whether eccentric orbits would improve constellation efficiency. 
Consider an eccentric orbit with apoapsis over the northern hemisphere: the reflector spends more time in view of the base and might have an improved duty cycle.
However, this would entail greater altitude of illumination, and likely greater slant distances, increasing spot size and thus reducing efficiency.
Eccentricity would also in general introduce secular drift in the argument of periapsis, which would limit the benefit of initial northern-hemisphere apoapsis. 
Critical-inclination orbits could~perhaps mitigate this.
Constellation design would become more complex, and the set of~permitted inclinations would change.
This merits further investigation.

\section{Conclusions}
\label{sec:conclusion}
\begin{enumerate}
    \item We show (via simulation) that a realistically agile low-Mars-orbit solar reflector can reflect light to a human base and station-keep using only solar radiation pressure and reaction-wheel-class three-axis control.
    \item We show (via simulation) that solar sails of areal density between 15–50 g/m$^2$ could fly from~low Earth orbit to low Mars orbit regardless of Earth--Mars phasing.
    \item For four families of Mars-warming orbits, we show how orbital parameters (altitude, local~time of the ascending node, initial mean anomaly) control energy delivery to a base (fraction~of day illuminated, delivery efficiency).
    \item We sketch a constellation design to double local year-averaged insolation.
\end{enumerate}

Open questions include how large individual reflectors would have to be in order to reduce the total number of spacecraft in Mars orbit to keep such a constellation operationally collision-safe.

We anticipate that progress toward doubling sunlight for a human Mars base would help drive progress in solar sail technology needed for more ambitious OSR architectures, such as those capable of sublimating Mars' buried south polar CO$_2$ ice [\cite{buhler_coevolution_2020}] to double atmospheric pressure in support of terraforming [\cite{kite_research_2026}]. 

\section*{Acknowledgments}
\label{sec:acknowledgements}
We thank Erika DeBenedictis, Charlie Garcia, and Yuri Shimane for discussions, and Erika DeBenedictis and Ashwin Braude for reading a draft.
Claude Code and Codex were used to develop the simulation software, run scripts, and improve plots.
The authors assume full and exclusive responsibility for all aspects of the work. 
This work was partly funded by Astera Institute.

\section*{Data Availability}
\label{sec:data-availability}
Code is available on GitHub (\url{https://github.com/ariessunfeld/mars-osr}) and Zenodo (\url{https://doi.org/10.5281/zenodo.22168117}).

\printbibliography

\appendix
\numberwithin{equation}{section}

\titleformat{\section}
{\large\sffamily\bfseries\color{paperblack}}
{Appendix~\thesection:}
{0.75em}
{}

\section{Reflected-spot geometry and delivered irradiance (\S\ref{methods:delivering-sunlight})}
\label{app:reflected-spot}

Given the~sail's position and attitude, the~Sun's position, and a surface base, how much sunlight (W/m$^2$) does the~sail reflect toward the~base, and over what ``spot'' area is the~reflected energy spread?
We follow \citet{canady_illumination_1982} and \citet{celik_analytical_2022} and work in the~Mars-centered J2000 frame.
The location of the~base is mapped to inertial coordinates through the~rotating Mars body-fixed frame (\texttt{IAU\_Mars}).
We use $\mathbf r$ and $\mathbf r_{\rm t}$ for the~sail and base positions, $d = \lVert \mathbf r_{\rm t} - \mathbf r\rVert$ for~the~sail$\to$base slant-range distance, $r_\odot$ for the~sail$\to$Sun distance, $\that$ for the~sail$\to$base unit vector ($\that$ here no longer refers to the~toward-velocity vector as it does in Eq.~\ref{eq:esc-normal}), and $\zhat$ for the~base's outward normal vector.

Our model accounts for the~distance to and finite angular size of the~Sun, the~oblique-projection ellipse, sail foreshortening, non-ideal specular reflectance, the~Mars umbra, and the~base's horizon, assuming a spherical Mars.
We omit atmospheric extinction (we set $\chi=1$); we take the~sail to be flat and rigid (no wrinkling); we do not model the~sail's pointing error; we ignore the~diffuse/off-specular lobe; and we do not account for local hills or slopes at the~target.

Seasonal variation is modest because the~$1/r^2$ dependence cancels: as the~Sun dims with distance, it also shrinks in angular size, so the~reflected solar image shrinks in the~same proportion and the~concentration rises to compensate exactly. 

\subsection{Optimal bisector pointing}
\label{app:reflected-spot:bisector-pointing}

A mirror reflects the~Sun onto the~target when its normal $\nhat$ bisects (see Section~\ref{disc:sensitivity-to-pointing-error}) the~sail$\to$Sun and sail$\to$base directions: ``optimal pointing'' (Section \ref{methods:delivering-sunlight}).
This orientation is
\begin{equation}
  \nhat^\star = \frac{\shat + \that}{\lVert \shat + \that\rVert},
  \qquad
  \cos\tfrac{\psi}{2} = \nhat^\star\cdot\shat = \nhat^\star\cdot\that,
  \label{eq:app-bisector}
\end{equation}
where $\psi$ is the~full Sun--sail--target angle and $\psi/2$ is the~angle of incidence at the~mirror.
We compute the~delivered irradiance assuming optimal pointing.

\subsection{Spot size and shape}
\label{app:reflected-spot:size-and-shape}

The reflected spot is an image of the~Sun's disk, because the~Sun has a finite angular diameter $\alpha = 2\arcsin(R_\odot / r_\odot)$, which diverges the~reflected beam.
On a plane perpendicular to the~beam at slant range $d$, the~image is a circle of radius $b = d\tan(\alpha/2)$.
Projected onto the~ground (which the~beam meets at elevation $\varepsilon$) the~circle becomes an ellipse with semi-minor axis $b$ (across the~ground trace of the~beam) and semi-major axis $a = b/\sin\varepsilon$ (along the~trace).
The sail's elevation $\varepsilon$ above the~target's horizon can be obtained via $\sin\varepsilon = (\mathbf r - \mathbf r_{\rm t})\cdot\zhat\,/\,d$.
This gives a spot area
\begin{equation}
  A_{\rm im} = \frac{\pi b^2}{\sin\varepsilon} = \frac{\pi\,[\,d\tan(\alpha/2)\,]^2}{\sin\varepsilon}
  \label{eq:app-spotarea}
\end{equation}
[\cite{canady_illumination_1982}, \cite{celik_analytical_2022}].
The spot is round when the~sail is directly overhead and grows long as the~sail approaches the~horizon.
This equation holds when the~sail diameter is much less than the~slant range [\cite{celik_analytical_2022}'s Eq.~15], which is the~case here.

\subsection{Delivered irradiance}
\label{app:reflected-spot:delivered-irradiance}

The mean irradiance over the~spot is the~reflected power spread over the~image area:
\begin{equation}
  I_{\rm t} = \frac{\eta\,\chi\,I_0(r_\odot)\,A\,\cos(\psi/2)\,\sin\varepsilon}{\pi\,[\,d\tan(\alpha/2)\,]^2},
  \qquad I_0(r_\odot) = \frac{L_\odot}{4\pi r_\odot^2},
  \label{eq:app-irradiance}
\end{equation}
where $A$ is the~sail area, $I_0$ is the~solar irradiance at the~sail, $\varepsilon$ is the~sail's angle of elevation from the~base's perspective (\citet{canady_illumination_1982}'s Eq.~9).
The numerator accounts for the~sail's orientation via $\cos(\psi/2)$, the~target's inclination via $\sin\varepsilon$, and the~sail's distance from the~Sun via $I_0\propto r_\odot^{-2}$.
The reflectance $\eta = \rho s$ is the~specular fraction of the~same non-ideal \citet{mcinnes_solar_1999} optical model we use for the~solar radiation pressure force, where $\rho$ is the~total reflectance and $s$ the~specular fraction; for the~aluminized square sail, $\eta = 0.83$.
Only this specular fraction forms a~directed~beam: the~diffuse fraction $\rho(1-s)$ and the~absorbed fraction $1-\rho$ scatter or re-radiate over $2\pi$~steradians and deliver negligible flux to the~target.
The factor $\chi$ is the~broadband direct-beam atmospheric transmission defined in Appendix~\ref{app:atmospheric-transmission}. 
We set $\chi=1$, ignoring atmospheric transmission losses, except when modeling OSR whole-constellation power-delivery and microclimate scenarios.

We define the~``delivered fluence'' at the~target over one pass as the~time integral $F = \int I_{\rm t}\,\mathrm{d}t$ [J/m$^2$] across the~delivery window, and sol-averaged fluence similarly. 
We report $I_{\rm t}$ as the~mean irradiance over the~spot (the total reflected power divided by the~image area).
That is, we assume the~spot lands on the~target and that the~spot is bigger than the~target.

The delivered irradiance is zero unless
(1) the~sail is sunlit,
(2) the~sail is above the~base's horizon ($\varepsilon \geq \varepsilon_{\min}$; we use $\varepsilon_{\min} = 10^\circ$, a horizon mask that excludes low passes where long slant paths and~possible near-base terrain make for poor delivery), and
(3) the~Sun-sail-target angle is $\leq168.5^\circ$ ($\cos(\psi/2)\geq 0.1$). 
The third condition excludes nearly antiparallel Sun and target directions, as~viewed from the~sail, which would require nearly edge-on-to-Sun reflector orientation.

\section{Atmospheric transmission}
\label{app:atmospheric-transmission}
We use the single-layer atmosphere model from \citet{vicente-retortillo_model_2015} to calculate atmospheric extinction (i.e., scattering and absorption). We thus obtain the surface irradiances for natural sunlight and for the reflected light, based on their respective top-of-atmosphere (TOA) values. 
Extinction by CO$_2$ gas molecules and water-ice clouds is small [\cite{vicente-retortillo_model_2015}], so we only consider dust opacity. 
Mars' dust cycle varies from year~to~year [\cite{montabone_eight-year_2015}]. 
For simplicity, we use an analytical parameterization for the climatological  (excluding major dust storms) zonally-averaged column opacity as a function of latitude and season. 
To obtain this parameterization (Eqs.~\ref{eq:tau-dust}), we fit the NASA Ames Global Climate Model (GCM) datafile \texttt{DustScenario\_Background.nc} available at \url{github.com/nasa/AmesGCM/blob/main/data/DustScenario_Background.nc}. 
The dust opacity $\tau$ is given by

\begin{equation}
        \tau(\phi,L_s)=\exp\left[\mathcal{T}(\phi,L_s)\right],
        \label{eq:tau-dust}
\end{equation}
where $\phi$ and $L_s$ are measured in degrees, and

\begin{align}
\mathcal{T}(\phi,L_s) ={}
-1.3950 & \nonumber 
 +0.8696\cos\!\left[\frac{\pi}{180}\left(L_s+142.5863\right)\right] \\ \nonumber   +\,0.1090\cos\!\left[\frac{\pi}{180}\left(3\phi-130.9271\right)\right] \nonumber 
 & +0.8918\cos\!\left[\frac{\pi}{180}\left(\phi-L_s+6.9158\right)\right] \\ \nonumber +\,0.5533\cos\!\left[\frac{\pi}{180}\left(2\phi+14.3980\right)\right]  \nonumber & +0.3711\cos\!\left[\frac{\pi}{180}\left(2\phi-L_s-141.5641\right)\right]
\label{eq:T-dust}
\end{align}

We use Planck-weighted (between 0.244-4.45$\mu {\rm m}$) averages from \cite{wolff_wavelength_2009} for the single-scattering albedo $w_0$ =  0.914 and $g$ =  0.724 to model the dust. 
With the column dust opacity $\tau$ constrained, \citet{vicente-retortillo_model_2015}'s model splits solar and the reflectors' irradiances into their direct ($\chi^{\rm DIR}$) and diffuse ($\chi^{\rm DIFF}$) components based on their respective solar-zenith angles $\mu_0^{\rm Sun}$ and $\mu_0^{\rm reflector}$. 
For the reflectors, we use a power-weighted average solar-zenith angle computed over the constellation at each timestep.

To calculate the total irradiance available at the surface, we include both the direct and diffuse (``all-sky'') components for the Sun, but we only retain the direct component for the reflectors. Due~to the relatively narrow (km-scale) beam from each reflector, most of the light scattered within that path is unlikely to hit the target location.

\begin{equation}
    I_{\rm total}^{\rm SFC}
    \;=\; 
    I_{\rm Sun}^{\rm TOA} 
    \cdot
    (\chi_{\rm Sun}^{\rm DIR} + \chi_{\rm Sun}^{\rm DIFF}) 
    \;+\;
    I_{\rm reflectors}^{\rm TOA} 
    \cdot 
    \chi_{\rm reflectors}^{\rm DIR}  
    \label{eq:I-sail-sun}
\end{equation}

\section{1D climate model}
\label{app:1D-climate-model}

The geothermal and sensible heat fluxes (minor contributions to the budget compared to radiative and conductive fluxes) are defined as in \citet{kling_subsistence_2020}. 
While the downward infrared flux at the surface can be rigorously obtained by using a radiative transfer model, for simplicity we instead parameterize $Q_{\rm IR \downarrow}$ as a function of the surface temperature by applying a quadratic fit to the $Q_{\rm IR \downarrow}$ values from a full Global Climate Model (GCM) simulation\footnote{available at \url{https://data.nas.nasa.gov/mcmc/portals/web-interface}} with climatological dust:

\begin{equation}
    \begin{aligned}
    Q_{IR \downarrow} &= A + BT_s+CT_s^2 \\
    A &= -2.63898123 \\
    B &= 0.145363974 \\
    C &= 8.36913082 \times 10^{-5},
    \end{aligned}
\end{equation}

where $T_s$ is the surface temperature. 
The coefficient of determination for the fit is $R^2=0.70$, yet this provides an adequate representation for the greenhouse warming by the atmosphere, including its~dependence on (surface) temperatures. 
We use the following values to represent soil conditions at 40$^\circ$ North: 

\begin{table}[H]
    \centering
    \begin{tabular}{lccc}
        \toprule
        Parameter & Symbol & Value & Units \\
        \midrule
        Thermal conductivity & $k$
            & $3\times10^{-2}$ & $~\mathrm{W\,m^{-1}\,K^{-1}}$ \\
        Regolith density & $\rho_{\mathrm{reg}}$
            & $1.481\times10^3$ & $~\mathrm{kg\,m^{-3}}$ \\
        Regolith specific heat capacity & $c_{p,\mathrm{reg}}$
            & $7.359\times10^2$ & $~\mathrm{J\,kg^{-1}\,K^{-1}}$ \\
        Albedo & $a$
            & $2.3\times 10^{-1}$ & --- \\
        Stefan--Boltzmann constant & $\sigma$
            & $5.67 \times 10^{-8}$ & $~\mathrm{W\,m^{-2}\,\mathrm{K}^{-4}}$ \\
        \bottomrule
    \end{tabular}
\end{table}

\section{Derivation of Earth-escape steering law}
\label{app:energy-rate-derivation}
The steering law of Section \ref{methods:earth-escape} maximizes $\dot\varepsilon$, where
$\varepsilon=\tfrac12\,\mathbf v\cdot\mathbf v-\mu_\oplus/r$ is the osculating
two-body energy (Eq.~\ref{eq:esc-energy}, $r=\lVert\mathbf r\rVert$). We first split
the equation of motion into the central term and~the~perturbing acceleration
$\mathbf a_{\mathrm p}$ (oblateness, third bodies, solar radiation pressure, drag):
\begin{equation}
  \dot{\mathbf v}=-\frac{\mu_\oplus}{r^{3}}\mathbf r+\mathbf a_{\mathrm p}.
  \label{eq:app-eom}
\end{equation}
With $\dot r=(\mathbf r\cdot\mathbf v)/r$, differentiating $\varepsilon$ and
substituting Eq.~\ref{eq:app-eom} gives
\begin{equation}
  \dot\varepsilon
  =\mathbf v\cdot\dot{\mathbf v}+\mu_\oplus\frac{\mathbf r\cdot\mathbf v}{r^{3}}
  =\Bigl(-\frac{\mu_\oplus}{r^{3}}\,\mathbf r\cdot\mathbf v
         +\mathbf v\cdot\mathbf a_{\mathrm p}\Bigr)
   +\mu_\oplus\frac{\mathbf r\cdot\mathbf v}{r^{3}}
  =\mathbf a_{\mathrm p}\cdot\mathbf v,
  \label{eq:app-edot-final}
\end{equation}
where the central terms have canceled. (This is consistent with the work--energy theorem for $\varepsilon$:
two-body gravity is conservative with respect to $\varepsilon$ and does no net work, so only $\mathbf a_{\mathrm p}$ changes it.)

Since $\mathbf a_{\mathrm p}=\mathbf a_{J_2}+\mathbf a_{3\mathrm b}
+\mathbf a_s+\mathbf a_d$, only the solar radiation pressure and drag terms $\mathbf a_s,\,\mathbf a_d$ depend
on~the~normal $\hat{\mathbf n}(\alpha)$ (Eq.~\ref{eq:esc-normal}). Therefore (as stated in Eq.~\ref{eq:esc-argmax}),
\begin{equation}
  \operatorname*{arg\,max}_{\lVert\alpha\rVert\le\alpha_{\mathrm c}}\dot\varepsilon
  =\operatorname*{arg\,max}_{\lVert\alpha\rVert\le\alpha_{\mathrm c}}
   (\mathbf a_s+\mathbf a_d)\cdot\mathbf v
  =\operatorname*{arg\,max}_{\lVert\alpha\rVert\le\alpha_{\mathrm c}}
   (\mathbf a_s+\mathbf a_d)\cdot\hat{\mathbf v},
  \label{eq:app-merit}
\end{equation}

\section{Interplanetary trajectory optimization}
\label{app:interplanetary-traj-opt}
For each Earth escape state, we seek a corresponding Mars capture state which can be reached in~$\lesssim3.5$~years from that Earth escape state.
We approach this as an optimization problem, where the optimizer's task is to find an attitude profile for~the~sail that produces a trajectory which delivers it to the Mars capture state.
We divide the attitude profile into sixteen segments, equal in duration, that span the full transfer duration $D$.
We~parameterize each segment by angles $\theta$ and $\phi$, where $\theta$~is~the~tilt out of the initial heliocentric orbit plane, and $\phi$ is~the~in-plane pitch (the angle between the Sun-line and the along-track direction).

To define these angles, let $\hat{\mathbf k}$ be the sail's heliocentric orbit normal at the start of the transfer.
At time $t$, let $\shat(t)$ be the sail-to-Sun unit vector and define the in-plane transverse direction $\hat{\mathbf q}(t)=\bigl(\hat{\mathbf k}\times\shat\bigr)/\lVert\hat{\mathbf k}\times\shat\rVert$.
The commanded sail normal during segment $i$, denoted by $\nhat_i(t)$, is
\begin{align}
    \nhat_i(t)
    &= \frac{\cos\theta_i\,\mathbf n_{{\rm p},i}(t)+\sin\theta_i\,\hat{\mathbf k}}
    {\left\lVert\cos\theta_i\,\mathbf n_{{\rm p},i}(t)+\sin\theta_i\,\hat{\mathbf k}\right\rVert}, \\
    \text{where}~~\mathbf n_{{\rm p},i}(t)
    &= \cos\phi_i\,\shat(t) + \sin\phi_i\,\hat{\mathbf q}(t).
    \label{eq:interplanetary-attitude}
\end{align}
We bound both angles as follows:
\begin{equation}
    -55^\circ\leq\phi_i\leq55^\circ,
    \qquad
    -55^\circ\leq\theta_i\leq55^\circ.
    \label{eq:interplanetary-angle-bounds}
\end{equation}
Each segment lasts tens of days, ample for each reorientation given the slew limits in Eq.~\ref{eq:esc-slew}.

Rather than fix the flight duration, we let the optimizer vary duration within the bounds $D_{\min} = t_{\rm cap} - t_{\rm esc}$ and  $D_{\max} = D_{\min} + 28$~days.
The targeted final state is thus not necessarily $X_{\rm cap}$ exactly, but rather a function of duration, looked up using the capture spiral's position and velocity (recorded as simulation output) and interpolated for continuity.

The bounds of 500~days and 1,300~days are empirical; for the areal densities considered, no $<500$-day interplanetary transfers  were found, and each Earth--Mars phasing considered permitted $<1{,}300$-day interplanetary transfers.
We use 28 days because for all of the Mars captures simulated, the capture spiral can be traced inward from its outermost point for $\ge$~28 days before undergoing one full revolution in the Mars-centered frame.
(Directly targeting a state >~1~revolution into the capture spiral is more difficult.)
Moreover, the relatively small value helps avoid confounding the optimizer with multiple minima.

To find such a trajectory, we first define the lookup function $f(X_{\rm cap},\,\,d)$ as the heliocentric state vector $(\mathbf r,\,\mathbf v)$ from $X_{\rm cap}$'s capture spiral at time $t_{\rm cap} + d$.
When $d=0$, $f(X_{\rm cap},\,\,0) = (\mathbf r_{\rm cap}, \mathbf v_{\rm cap})$.
With $f$ thus defined, we write the decision vector as
\begin{equation}
    \mathbf c
    =
    [\phi_0,\ldots,\phi_{N},\,
    \theta_0,\ldots,\theta_{N},\,D].
    \label{eq:interplanetary-decision-vector}
\end{equation}
Propagating the reflector for $D$ days gives the final state $(\mathbf r_{\rm final},\mathbf v_{\rm final})$ at $t_{\rm esc}+D$.
We use $N=15$ and employ fourth-order Runge--Kutta with uniform steps of at most 7,200~s to propagate.
The corresponding target state is
\begin{equation}
    (\mathbf r_{\rm target},\mathbf v_{\rm target})
    =
    f\!\left(
    X_{\rm cap},\,
    D-(t_{\rm cap}-t_{\rm esc})
    \right).
\end{equation}

We normalize the error metric $\mathbf g(\mathbf c)$ as follows:
\begin{equation}
    \mathbf g(\mathbf c)
    =
    \begin{bmatrix}
        \bigl(\mathbf r_{\rm final}-\mathbf r_{\rm target}\bigr)/r_{\rm s} \\
        \bigl(\mathbf v_{\rm final}-\mathbf v_{\rm target}\bigr)/v_{\rm s}
    \end{bmatrix},
    \qquad
    r_{\rm s}=1~{\rm AU},
    \qquad
    v_{\rm s}=\frac{\lVert\mathbf v_\oplus(t_{\rm esc})\rVert}{2\pi}.
    \label{eq:interplanetary-defect}
\end{equation}
We do this so that the optimizer gives position error and velocity error comparable numerical weight, despite the different units involved.

We then solve each candidate transfer using Interior Point Optimization (IPOPT) through its~Python interface, \texttt{cyipopt}.
The solve firstly finds a nearby solution from cold start but doesn't necessarily reduce the position and velocity errors to zero, and secondly tries to refine the first-stage solution. The first part is a bounds-constrained least-squares problem:
\begin{equation}
    \underset{\mathbf c}{\operatorname{minimize}}
    \quad
    J_{\rm LS}(\mathbf c)
    =
    \lVert\mathbf g(\mathbf c)\rVert_2^2,
    \label{eq:interplanetary-least-squares-cost}
\end{equation}
subject to Eq.~\ref{eq:interplanetary-angle-bounds} and $D\in[D_{\min},D_{\max}]$.

This starts from a constant-angle initial guess with $D$ at the center of its 28-day interval. The second part starts from the best solution found in the first stage and solves
\begin{equation}
    \underset{\mathbf c}{\operatorname{minimize}}
    \quad 0
    \qquad
    \text{subject to}
    \qquad
    \mathbf g(\mathbf c)=\mathbf 0,
    \label{eq:interplanetary-feasibility-polish}
\end{equation}
with the same bounds.
Although the objective in Eq.~\ref{eq:interplanetary-feasibility-polish} is zero, IPOPT further reduces infeasibility associated with the six equality constraints $\mathbf g(\mathbf c)$.
We set the equality-constraint violation tolerance in the second part to $10^{-8}$ and allow up to 300 iterations.

\section{Low Mars orbit optimization}
\label{app:low-mars-orbit-optimization}
To optimize the attitude profile each sol for station-keeping and delivery, we use Python package SciPy's Differential Evolution (DE) global optimizer \citep{storn_differential_1997} with the following parameters:
population size = 15, initialization = Sobol, seed = 42, generations = 30.
Following DE, we refine the solution using the Limited-memory Broyden-Fletcher-Goldfarb-Shanno with Bounds (L-BFGS-B) algorithm  [\cite{zhu_algorithm_1997}, \cite{byrd_limited_1995}].

At each optimizer iteration, we~parameterize the sail's attitude profile using truncated Fourier series (Eqs. \ref{eq:alpha} and \ref{eq:delta}), the coefficients of which are chosen by the optimizer within specified bounds.
Each set $\mathbf{c}$ of coefficients defines a distinct attitude profile for the sail.
The bounds on the parameters and the formulation of the attitude profile and the slews ensure that the finite-agility constraints (Eq.~\ref{eq:esc-slew}) are satisfied by the attitude profile.

Delivery windows are found as follows:
Once the optimizer has proposed a candidate station-keeping-only (``cruise'') attitude profile $\nhat^\dagger(u)$, the sail is first propagated according to this profile.
During propagation, the intervals of human-base visibility are recorded.
After propagation, the cruise attitude profile is replaced by a composite profile in which illumination-optimal (``delivery'') pointing is imposed during the previously-identified intervals, and slews are imposed adjacent to each delivery-pointing arc to avoid discontinuities between cruise and delivery.
The resulting composite profile $\nhat(u)$ is then propagated.
But because the delivery pointing and slews introduce previously unaccounted-for solar radiation pressure perturbation, the delivery windows must be re-identified on the new trajectory and compared to the previous estimate.
Delivery-window boundaries are thus updated.
This process (i.e., damped fixed-point iteration) repeats until the illumination-start times between iteration $n-1$ and iteration $n$ are stable to within 1~s and no windows have appeared or disappeared since the last iteration.
At convergence, the scheduled delivery windows coincide with actually achievable delivery opportunities.

\subsection{Attitude profile parameterization}
\label{methods:LMO:attitude-profile-parameterization}

The sail's cruise attitude profile $\nhat^\dagger(u)$ provides a stable orbit despite perturbations; the purpose of~the composite attitude profile $\nhat(u)$ is to provide both orbital stability and human-base illumination.
The~main perturbation is solar radiation pressure.
We let the cone and clock angles of Eq.~\ref{eq:normal} vary with the~argument of latitude $u$ (the in-plane angular position of the sail measured from the ascending node), rather than with time.
Thus, a single attitude profile applies on every orbit (with the exception of delivery windows and corresponding slews taking precedence during parts of some orbits).
We define the parameter space for control by representing the cone angle $\alpha$ as a~two-harmonic modulation about its mean value $\alpha_0$, while the clock angle $\delta$ advances once per orbit with argument of latitude $u$ and carries its own two-harmonic modulation about an offset $\delta_0$:
\begin{align}
\alpha(u) &= \alpha_0
  + \sum_{k=1}^{2}\bigl[a_{ck}\cos(k u) + a_{sk}\sin(k u)\bigr],
  \label{eq:alpha}\\[4pt]
\delta(u) &= u + \delta_0
  + \sum_{k=1}^{2}\bigl[d_{ck}\cos(k u) + d_{sk}\sin(k u)\bigr].
  \label{eq:delta}
\end{align}
Substituting $\alpha(u)$ and $\delta(u)$ into Eq.~\ref{eq:normal} provides the commanded normal at every point along the orbit. The design variables are defined as
\begin{equation}
\mathbf{c} = [\,\alpha_0,\,a_{c1},\,a_{s1},\,a_{c2},\,a_{s2},\;
                \delta_0,\,d_{c1},\,d_{s1},\,d_{c2},\,d_{s2}\,] \in \mathbb{R}^{10}.
\end{equation}

\subsection{Cost function}
\label{app:LMO_opt:cost-function}

Let $\mathbf{c}$ be the attitude-profile coefficients and let $\mathbf x(t)=(\mathbf r,\, \mathbf v)$ solve the equations of motion $\dot{\mathbf x} = f\bigl(\mathbf x,\, \nhat(t;\,\mathbf{c}),\, t\bigr)$ over $[t_0, t_1]$. We seek to minimize

\begin{equation}
    \label{cost-func-orbital-stability}
    \mathbf{J}(\mathbf{c}) = \frac{||\Delta \mathbf r_{\rm fix}||^2}{\sigma_r^2} + \frac{||\Delta \mathbf v_{\rm fix}||^2}{\sigma_v^2} + \frac{e_{\rm max}^2}{\sigma_e^2} + \lambda \cdot \bigl({\rm max}(0, F_{\rm floor} - F)\bigr)^2
\end{equation}

where $\Delta \mathbf r_{\rm fix},\,\,\Delta \mathbf v_{\rm fix}$ are the body-fixed end-minus-start deltas, $e_{\rm max}$ is the peak osculating eccentricity over the sol, $F$ is the delivered fluence (J/m$^2$ measured with a representative $1{,}000{\rm m}^2$ reflector), and ($\sigma_r$, $\sigma_v$, $\sigma_e$, $\lambda$)~=~(1000~km, 0.8~km/s, 10, 1) are weights. 
$F_{\rm floor} = 0.9\,F_{\rm opt}$, where $F_{\rm opt}$ is the~bisector-pointing optimal fluence that can be achieved in this orbit.
For multi-sol propagation, we additionally penalize the cumulative Mars-fixed position and velocity errors at the end of each sol relative to the initial state, using quadratic terms analogous to the first two terms in Eq.~\ref{cost-func-orbital-stability}.

\subsection{Numerical propagation in Mars orbit}
\label{app:LMO_opt:numerical-propagation}

For each candidate attitude profile, we integrate the Mars-centered J2000 equations of motion using SciPy's adaptive \texttt{DOP853} solver [\cite{hairer_solving_1993}] with tolerances $\mathrm{rtol}=10^{-9}$ and $\mathrm{atol}=10^{-6}$. 
Each integration spans one Mars solar day and returns an array of states on a 60~s grid, including the exact sol end state. This grid is used to identify delivery windows and calculate fluence. 
\texttt{DOP853} selects step size adaptively; 60~s is only an evaluation interval. 
The final state of each sol becomes the initial state for the next sol's attitude profile, which is independently re-optimized.
If a~new~initial~state is found during re-optimization, a slew respecting the constraints of Eq.~\ref{eq:esc-slew} is imposed between the final state of the previous sol and the initial state of the following sol.

\section{Figures with details of orbital requirements and performance}
\label{app:additional-figures}
\begin{figure}[H]
    \centering
    \includegraphics[width=0.6\textwidth]{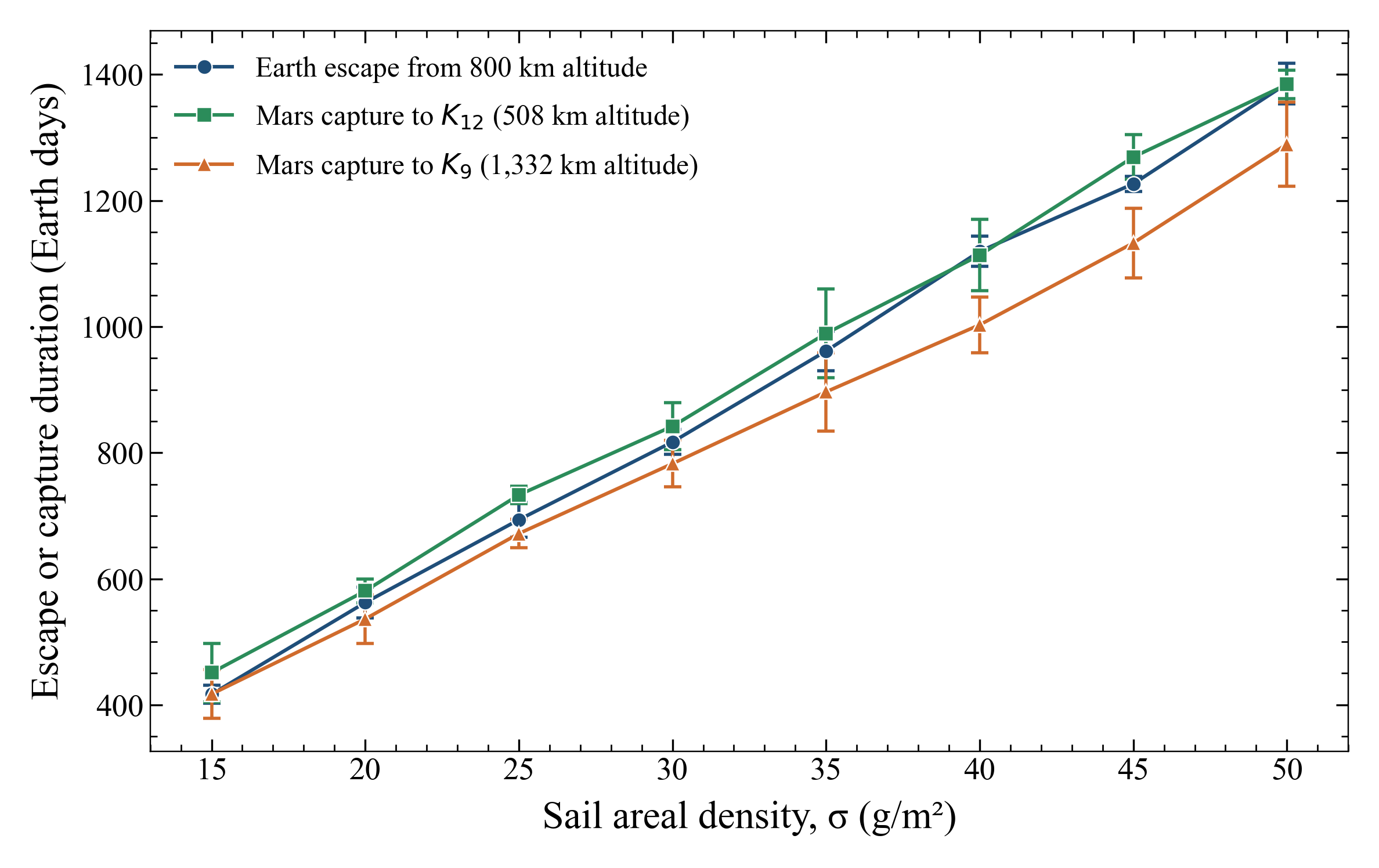}
    \caption{
    Escape duration and capture duration as a function of orbiting solar reflector areal density $\sigma$ (g/m$^2$). Points show mean durations among successful runs; whiskers show $\pm$1 standard deviation across successful runs. Mars capture samples six ($n=6$) arrival seasons separated by $60^\circ$ in solar longitude. All Mars points use $n=6$ except for $K_{12}$ at $\sigma=50$~g/m$^2$ which uses $n=4$.}
    \label{fig:time_vs_sigma}
\end{figure}

\begin{figure}[H]
    \centering
    \includegraphics[width=0.9\textwidth]{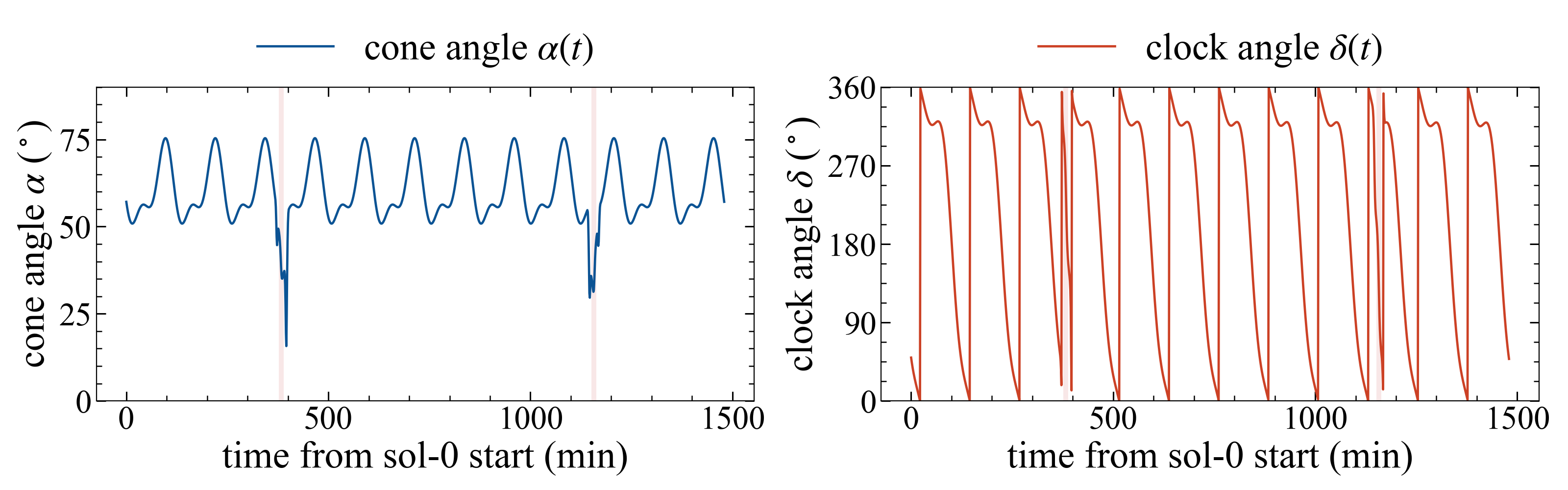}
    \caption{
    Time series of cone angle $\alpha$ and clock angle $\delta$ over one sol at Mars perihelion phasing for an 18~g/m$^2$ reflector in a near-polar, circular, retrograde, Sun-synchronous, $L_0$~=~18~h, $M_0=0$, repeat-ground-track orbit at 508~km altitude. 
    Pink bands and corresponding attitude spikes near 400 minutes and 1200 minutes correspond to delivery windows. 
    Closure after one sol is 0.6~km position error and 1.18~m/s velocity error.}
    \label{fig:alpha-and-delta-one-sol}
\end{figure}

\begin{figure}[H]
    \centering \includegraphics[width=0.95\textwidth]{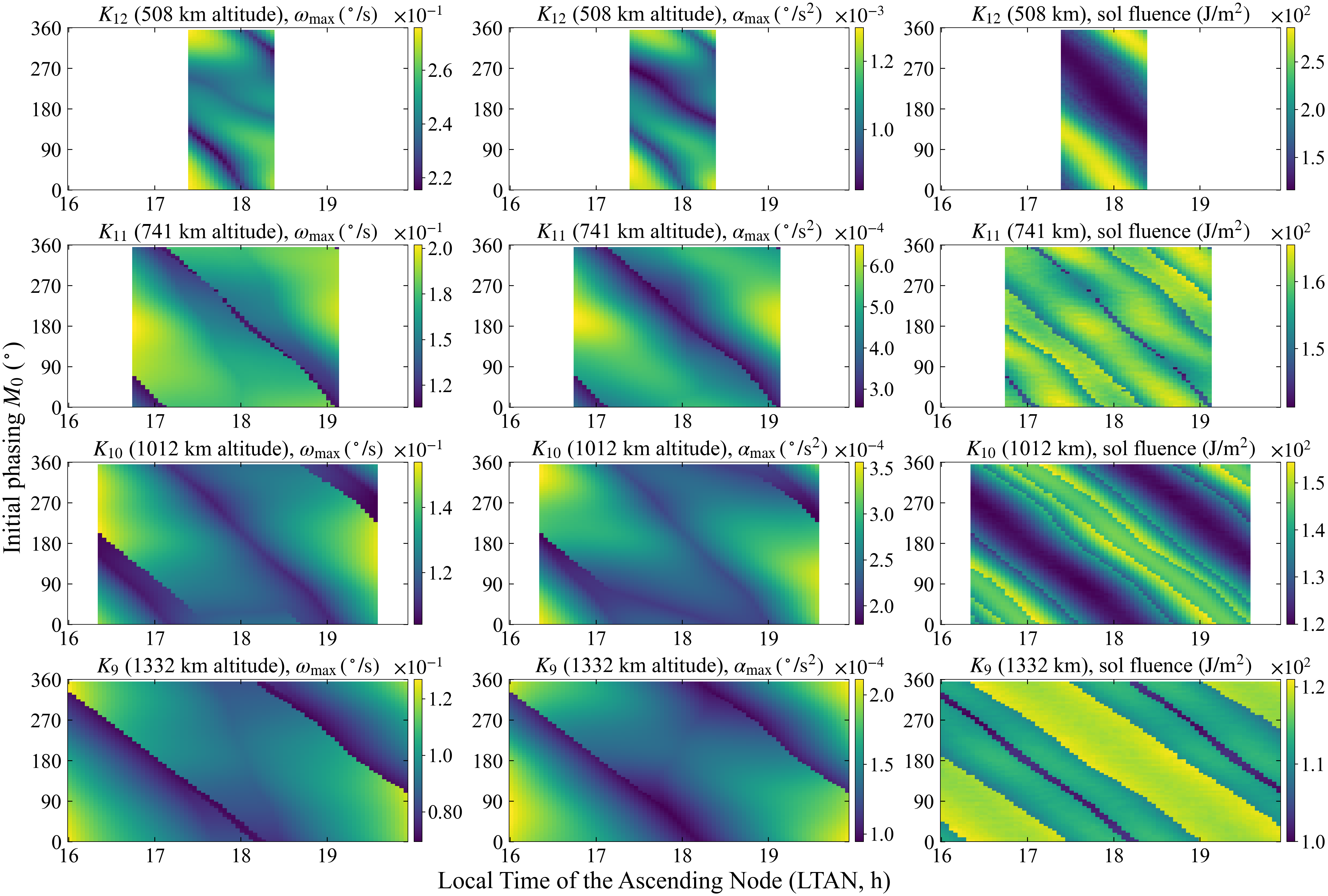}
    \caption{Peak slew demand (angular velocity and angular acceleration) and sol-averaged irradiance as~functions of LTAN and initial~mean anomaly $M_0$ for each of the four repeat-ground-track orbital altitudes studied.
    \textbf{Left column:} peak angular velocity during optimal pointing $(^\circ/{\rm s})$.
    \textbf{Middle column}: peak angular acceleration during optimal pointing $(^\circ/{\rm s^2})$.
    \textbf{Right column:} sol-integrated fluence (J/m$^2$ from a 10,000~m$^2$ reflector).
    \textbf{Top~row:} the $K_{12}$ (508~km altitude) orbits.
    \textbf{Second~row:} the $K_{11}$ orbits.
    \textbf{Third~row:}~the~$K_{10}$~orbits.
    \textbf{Bottom~row:} the $K_{9}$ orbits.
    \textbf{X~axes:} Local Time of the Ascending Node (LTAN, h).
    \textbf{Y~axes:}~Initial~mean~anomaly $M_0$.
    Colorbar range varies with each plot.
    }
    \label{fig:slew-demand}
\end{figure}

\begin{figure}[H]
    \centering
    \includegraphics[width=0.6\textwidth]{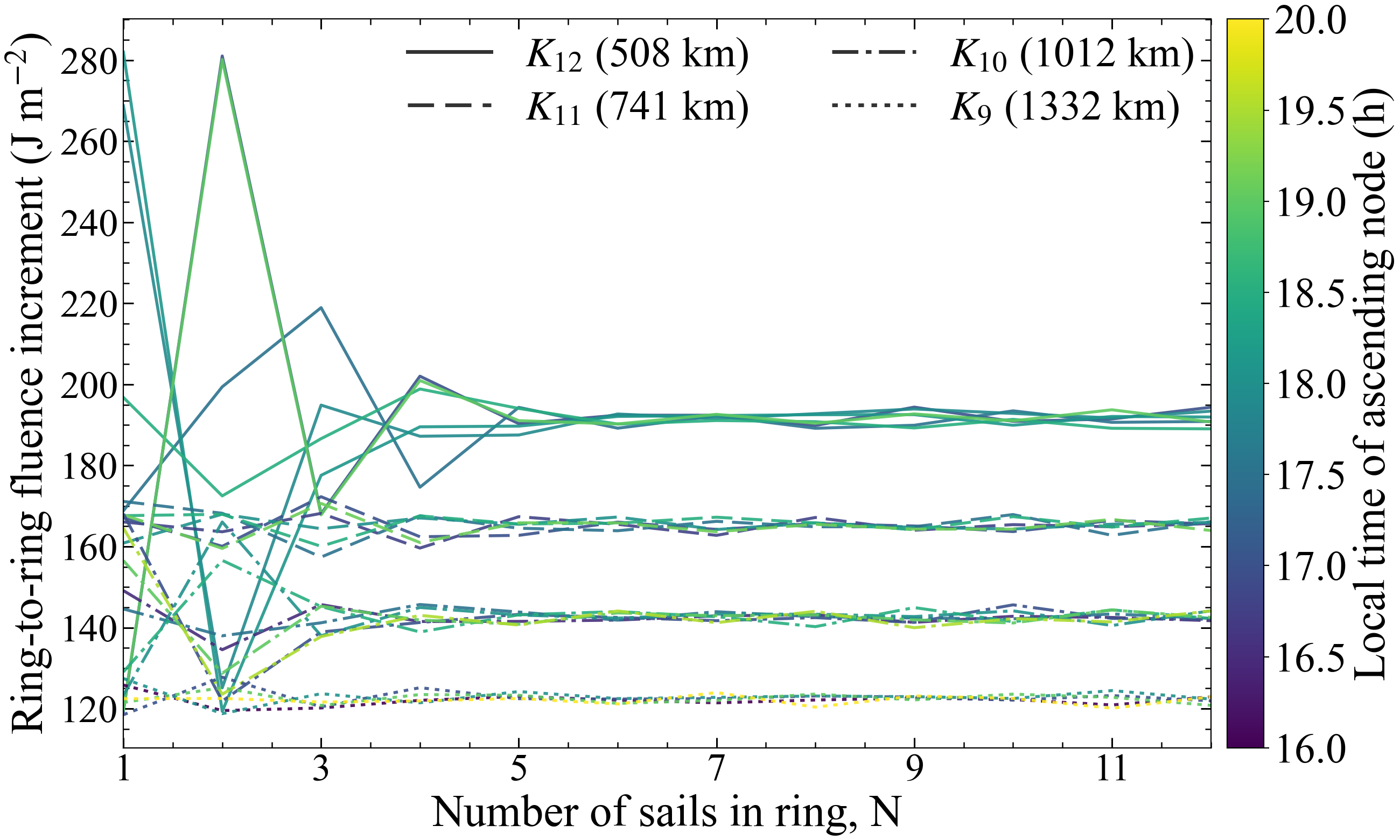}
    \caption{
    Incremental fluence (J/m$^2$) for a ring with $N$ equally spaced orbiting solar reflectors (OSRs) versus a ring with $N-1$ equally spaced OSRs, as a function of $N$. 
    Different line styles denote different orbital altitudes. 
    From top to bottom are the $K_{12}$, $K_{11}$, $K_{10}$, and $K_9$ orbital altitudes. 
    After $N\approx6$ the marginal fluence becomes approximately constant (insensitive to phasing).}
    \label{fig:ring-to-ring-fluence}
\end{figure}

\section{Benchmarking and verification}
\label{app:benchmarking-and-verification}
\subsection{Mars gravity}
\label{app:bench:mars-gravity}

The Mars gravity calculation reads the~normalized \texttt{MRO120F} coefficients and their 3396.0~km \texttt{MRO120F} reference radius [\cite{konopliv_detection_2020}]. 
The normalized $\bar C_{2,0}$, $\bar C_{2,1}$, $\bar S_{2,1}$, $\bar C_{2,2}$, and $\bar S_{2,2}$ coefficients are also compared with the~\texttt{MRO120F} file. 

We evaluate zonal and tesseral accelerations against the~recurrences of \citet{cunningham_computation_1970}'s Eqs.~14--17. 
With only the~central term retained, the~result agrees with $-\mu\mathbf r/r^3$ to relative error below $10^{-14}$. 
With order zero retained through degrees 2, 4, and 10, the~Cunningham result agrees with a~separately implemented scalar Legendre recurrence to relative error below $10^{-13}$ at 100 random positions per degree. 
At degree two, the~scalar recurrence also agrees with the~$J_2$ acceleration to relative error below $10^{-13}$. 
We include tests that evaluate the~$C_{2,2}$, $S_{2,2}$, and $C_{2,1}$ accelerations at the~equatorial prime meridian to check that the~sign of each component is~correct.

For a $J_2$-only Mars model, we propagate test particles at inclinations $30^\circ$, $60^\circ$, and $80^\circ$ and fit the~secular nodal rates. 
These agree with the~first-order analytical rate of \citet{brouwer_solution_1959} to within 1\% over 30 orbits. 
The fitted argument-of-periapsis rate for an orbit with $e=0.05$ and $i=45^\circ$ agrees to within 2\%. 
In three- to five-orbit tests, relative total-energy drift remains below $10^{-9}$ and fixed-axis axial-angular-momentum drift below $10^{-8}$ in the~zonal model.
Tesseral harmonics produce a measurable loss of axial symmetry.

The Mars-year station-keeping calculation used degree and order six. 
We replayed the~saved attitude histories for sols 1, 334, and 668. 
Tightening the~degree-six DOP853 tolerances from ($\mathrm{rtol}=10^{-9}$,~$\mathrm{atol}=10^{-6}$) to ($10^{-13}$, $10^{-12}$) changed position by only $\sim$10~m over one sol. 

All body-fixed gravity evaluations use the~SPICE transformation from J2000 to \texttt{IAU\_MARS} at the~current ephemeris time and rotate the~acceleration back to J2000. 
Mars capture runs the~ephemeris clock backward. 
A forward-then-reversed propagation with Mars gravity and solar third-body gravity retraces the~initial state to within $10^{-4}$~km and $10^{-7}$~km/s.

\subsection{Solar radiation pressure and reflected-beam optics}
\label{app:bench:solar-rad-pressure}

The flat-sail acceleration is evaluated against limiting cases of the~non-ideal optical model in \citet{mcinnes_solar_1999}'s~Eq.~2.57. 
A face-on ideal mirror, pure absorber, and perfect Lambertian reflector produce $2PA/m$, $PA/m$, and $5PA/(3m)$, respectively, to relative error below $10^{-12}$. 
An edge-on sail and a sail in umbra produce zero acceleration. 
We test for and obtain linear scaling with area, inverse scaling with mass, inverse-square scaling with sail--Sun distance, and the~$\cos\alpha$ and $\cos^2\alpha$ incidence-angle relationships for a pure absorber and an ideal specular reflector, respectively. 

The reflected-spot calculation's solar-image semiaxes reproduce \citet{canady_illumination_1982}'s Eqs.~1--2 and the~image-area equations of \citet{celik_analytical_2022}'s Eqs.~8 and~13 to relative error below $10^{-14}$. 
Dividing reflected power by that area agrees to relative error below $10^{-5}$ with a~radiance calculation based on conservation of etendue [\cite{born_chapter_1980}'s \S4.8.3, Eqs.~22 and~25].

For a 1,000~m$^2$ reflector at 500~km slant range, at zenith and face-on, the~calculation gives 0.065--0.068~W/m$^2$ in vacuum and a perpendicular solar image radius of 1.6--1.7~km. 
Retaining the~finite mirror diameter term changes the~irradiance by less than 0.1\% at this scale. 
Our implementation rejects sail size and altitude configurations for which the~reflector is too large relative to the~reflected spot for the~point-reflector approximation.

\subsection{Circum-martian dust benchmark}
\label{app:bench:circum-martian-dust}

\citet{hamilton_circumplanetary_1996} model the~eccentricity of dust grains ejected from Mars' moons Phobos and Deimos. 
To check central gravity, $J_2$, solar radiation pressure, osculating elements, and long propagation, we compare with their orbit-averaged theory. To do this, we truncate our model to their assumptions: 
a circular and coplanar Mars orbit, parallel solar rays, no eclipse, spherical dust grains, and only Mars central gravity, $J_2$, and solar radiation pressure acting on the~dust grains.
We directly integrate their two-dimensional orbit-averaged equations, and we separately perform Cartesian integration using the~same reduced forces.

The orbit-averaged integration reproduces the~Deimos eccentricity expression (\citet{hamilton_circumplanetary_1996}'s Eq.~17). 
For Phobos, the~maximum-eccentricity jump is at a grain radius of approximately 331.5~$\mu$m and $e_{\max}\simeq0.25$ (\citet{hamilton_circumplanetary_1996}'s Fig.~2), and the~stationary-point bifurcation is near 232~$\mu$m and $e\simeq0.180$ (\citet{hamilton_circumplanetary_1996}'s Fig.~8).
The Cartesian Deimos maxima agree with the~orbit-averaged results to 0.001--0.017\% over radii 20--1000~$\mu$m. 
For Phobos, the~differences are 0.012--0.066\% in magnitude at 100--200~$\mu$m and 1.66--2.12\% at 400--1000~$\mu$m. 
We exclude grain sizes between 200--400$\mu$m from the~Phobos comparison because the~orbit-averaged period diverges there. 
Tightening the~Cartesian tolerances, doubling the~samples per orbit, and halving the~maximum step changes the~sampled maxima by $<0.0051\%$.

\subsection{Trajectory and delivery window timestep resolution}
\label{app:bench:trajectory-delivery-resolution}

We check the~Earth-escape drag acceleration against the~circular-orbit decay rate of \citet{king-hele_theory_1964}'s Ch.~4, \S18 (esp. Eq.~4.84--4.85). 
At a circular-orbit test state, we convert the~numerically evaluated drag acceleration to its implied semimajor-axis decay rate $\dot a=(2a^2/\mu)\mathbf v\cdot\mathbf a_{\rm d}$.
This agrees with the~analytical King-Hele decay rate to relative error below (10$^{-9}$).
A separate three-orbit integration with a face-on sail and exponential atmosphere agrees with the~analytical mean decay rate to within 3\%.

The heliocentric transfer shown in Fig.~\ref{fig:transfer} lasts $\approx1{,}138$~days. 
We replayed it with successive step halvings to check time-step convergence. 
This gave final-position convergence orders of 3.996, 3.998, and 4.270, consistent with the~expected fourth-order convergence of RK4.
Refining the~7200~s step ceiling to 1800~s changed the~final position by 7.293~m and the~final velocity by $1.395\times10^{-6}$~m/s. 
The final position and velocity errors in the~run shown in Fig.~\ref{fig:transfer} were 0.0366~km and $6.26\times10^{-5}$~m/s, well below the~10~km and 1~m/s limits considered as thresholds for ``closed'' by our implementation.

We also vary the~output cadence used to identify delivery windows (nominally 60~s, varied between 1~s and 240~s) and integrate fluence on five sols distributed evenly across one Mars year using the~optimized $K_{12}$ full-year trajectory (Table~\ref{tab:app-cadence-convergence}). 

\begin{table}[!htp]
\centering\small
\caption{
Delivery-window and fluence convergence for five $K_{12}$ sols. 
The trajectory and controller are fixed; only the~observation cadence is changed.}
\label{tab:app-cadence-convergence}
\begin{tabular}{rrrrr}
\toprule
Sol & 60~s fluence & 1~s fluence & 60~s error & Windows \\
 & (J/m$^{2}$) & (J/m$^{2}$) & (\%) & (60~s / 1~s) \\
\midrule
1   & 28.0834 & 28.3166 & $-0.824$ & 2 / 2 \\
168 & 26.5459 & 26.6509 & $-0.394$ & 2 / 2 \\
334 & 22.9522 & 23.1591 & $-0.893$ & 2 / 2 \\
501 & 24.4108 & 24.6270 & $-0.878$ & 2 / 2 \\
668 & 27.9237 & 28.2064 & $-1.002$ & 2 / 3 \\
\bottomrule
\end{tabular}
\end{table}

The 60~s calculation underestimates the~1~s fluence by 0.39--1.00\% in these five cases and recovers both delivery windows on every sol. 
The additional 1~s window on sol 668 is a short 181~s pass with peak elevation $10.579^\circ$, just above the~$10^\circ$ limit. 
It contributes 0.1000~J/m$^{2}$, or 0.355\% of that sol's 1~s fluence. 
A 120~s grid gives errors of 1.22--2.59\%. \
A 240~s grid misses one of the~primary windows in each sampled season. 
Thus the~nominal cadence slightly underestimates
fluence and can~omit short, grazing delivery windows while still resolving primary delivery windows.

\section{Constellation sizing}
\label{app:constellation-sizing}
We estimate how many OSRs could occupy the~continuously sunlit, Sun-synchronous altitude range identified in Section~\ref{res:families-of-orbits-for-MMaaS}.
Our estimate assumes circular orbits for the~OSRs in the~constellation and implicitly assumes that all OSRs can station-keep their circular orbits even more tightly than is~reported in Section~\ref{res:energy-and-stability:stability}.
Future work may reveal that a constellation consisting of fewer, individually larger, reflectors may be required to ensure collision-safe operation.

One altitude level is~called a ``shell,'' while one orbital plane within a shell is~called a ``ring.''
We consider circular shells from 508 to 1432~km altitude and include the~$K_{12}$, $K_{11}$, $K_{10}$, and $K_9$ repeat-ground-track altitudes of 508, 741, 1012, and 1332~km exactly.
Beginning at each reference altitude, shell centers advance in 5~km increments up to the~next reference altitude, which is~included exactly.
Above 1332~km, the~sequence continues in 5~km increments to 1432~km.
The resulting grid contains 185 shells, with adjacent separations of 5, 6, or 8~km.

$K_{\rm ref}$ denotes the~repeat-ground-track solution whose altitude is~nearest to a shell (Table~\ref{tab:constellation-sizing-summary}). 
The~reflectors in shells with altitudes not equal to repeat-ground-track altitudes do not follow exact repeat-ground-track orbits, but this does not prevent the~rings that they comprise from delivering sunlight to the~base.
We set the~Sun-synchronous inclination separately at each altitude using the~same first-order $J_2$ condition as in Section~\ref{methods:delivering-sunlight:constraints-from-orbit-assumptions}.

We compute the~whole-year eclipse-free LTAN band at every altitude by sampling 60 seasons over one Mars sidereal year, and at each season and trial LTAN, propagating a test particle in a circular orbit for 1.05 periods under central gravity, $J_2$, and solar third-body gravity. 
This trajectory is~sampled every 20~s and tested for full umbra.
Each seasonal interval is~shifted from true-Sun to mean-Sun LTAN using the~Mars equation of center; the~60 seasonal intervals are intersected, and the~widest subinterval centered on LTAN~=~18~h is~used.

\subsection{Geometric spacing and phasing method}
\label{app:constellation-sizing:method}

We impose a minimum 5~km separation between shell centers and a maximum plane density of one~ring per $5^\circ$ of right ascension of the~ascending node (RAAN).
Recent Starlink Gen2 authorizations\footnote{https://docs.fcc.gov/public/attachments/{DA}-26-36A1.pdf} use 5~km altitude increments and include shells with as many as 72 planes, corresponding to~$5^\circ$~spacing when distributed over $360^\circ$ of RAAN.
Starlink is~the~largest existing satellite constellation but an imperfect analogy in several ways: Starlink satellites are propulsive, have individually small collisional cross sections, are fewer in number than the~constellation we consider, and operate in~a~drag-cleaned environment.
The reflectors in the~constellation that we consider are individually less maneuverable, individually larger, greater in number by an order of magnitude than the~current number of Starlink satellites in orbit, and operate in extremely low-drag altitudes over Mars.

We impose a 300~km same-ring chord spacing and a 50~km closest approach between rings in~the~same shell:
\begin{equation}
    \Delta h = 5~\mathrm{km}, \qquad
    \Delta\Omega_{\min}=5^\circ, \qquad
    d_{\parallel}\geq300~\mathrm{km}, \qquad
    d_{\times}\geq50~\mathrm{km}.
    \label{eq:constellation-spacing}
\end{equation}

If the~eclipse-free interval spans $\Delta L$ hours, its RAAN span is~$\Delta\Omega=15^\circ\Delta L$ and the~number of rings per shell is~bounded by
\begin{equation}
    P \leq 1+\left\lfloor\frac{\Delta\Omega}{\Delta\Omega_{\min}}\right\rfloor .
    \label{eq:constellation-plane-count}
\end{equation}
For a candidate configuration of a given shell with $P$ rings and $S$ reflectors per ring, we distribute the~rings uniformly across the~LTAN interval and phase the~reflectors as
\begin{equation}
    u_{p,s}(t_0)
    =
    u_0+\frac{2\pi s}{S}+\frac{2\pi Fp}{PS},
    \qquad
    p=0,\ldots,P-1,\quad s=0,\ldots,S-1,
    \label{eq:constellation-walker}
\end{equation}
where $F\in\{0,\ldots,P-1\}$ is~the~Walker phasing parameter \citep{walker_satellite_1984}, and $u_{p,s}(t_0)$ is~the~initial argument of latitude of reflector $s$ in ring $p$.
For these circular orbits, initial mean anomaly $M_0~=~u_{p,s}(t_0)$. 
The phase increment between corresponding reflectors in neighboring rings is~$360^\circ F/(PS)$.
Changing $F$ shifts the~times at which reflectors pass through the~intersections of the~orbital planes but leaves the~ring locations and same-ring spacing unchanged.
The same-ring chord spacing is
\begin{equation}
    d_{\parallel}=2a\sin\left(\frac{\pi}{S}\right),
    \label{eq:constellation-along-track}
\end{equation}
for shell semimajor axis $a$.
It bounds the~number of reflectors in one ring according to
\begin{equation}
    S\leq S_{\max}(a)
    =\left\lfloor
    \frac{\pi}{\arcsin\!\left[d_{\parallel,\min}/(2a)\right]}
    \right\rfloor,
    \qquad d_{\parallel,\min}=300~\mathrm{km}.
    \label{eq:constellation-satellite-count}
\end{equation}

For each pair of rings, we require the~minimum Euclidean separation between their satellites over one circular orbit to exceed 50~km.
At each shell, we enumerate the~integers $P=1,\ldots,P_{\max}$, $S=2,\ldots,S_{\max}$, and $F=0,\ldots,P-1$.
We seek the~largest number of reflectors in that shell,
\begin{equation}
    (P^\star,S^\star,F^\star)
    =
    \underset{P,S,F}{\operatorname{arg\,max}}\;PS,
    \label{eq:constellation-integer-objective}
\end{equation}
subject to Eqs.~\ref{eq:constellation-spacing}, \ref{eq:constellation-plane-count}, and \ref{eq:constellation-satellite-count}.
When several feasible triples have the~same $PS$, we choose the~one with greater inter-ring clearance.
We resolve remaining exact ties by choosing smaller $P$ and then smaller $F$.

For instance, consider the~shell at 508~km.
Equations~\ref{eq:constellation-plane-count} and \ref{eq:constellation-satellite-count} give $P_{\max}=3$ and $S_{\max}=81$.
At~$P=3$ and $S=81$, $F=0$ gives a 24.70~km inter-ring minimum and fails the~screening criterion.
Both $F=1$ and $F=2$ are feasible, with minima of 76.09 and 50.99~km, respectively, so the~clearance tie-breaker selects $F=1$.
At 741~km, maximizing $PS$ selects $P=8$ and $S=55$, or 440 reflectors, and $F=4$ is~the~only phasing that reaches the~50~km inter-ring minimum.
At 1332~km, the~wider LTAN band permits $P=12$, but the~best feasible configuration with $P=12$ has $S=50$ and only 600 reflectors.
The selected $P=7$, $S=91$, $F=6$ configuration has 637 reflectors.
The minimum selected same-ring spacing is~300.231~km, and the~minimum selected equal-altitude inter-ring distance is~50.0036~km.

\subsection{Resulting representative constellation}
\label{app:constellation-sizing:result}

Table~\ref{tab:constellation-sizing-summary} gives the~constellation that results from the~approach described in Section~\ref{app:constellation-sizing:method}.
The number of rings and reflectors per ring changes with altitude, so we report ranges within each $K_{\rm ref}$ group.
These counts describe the~integer optima of our geometric constellation-sizing model, which assumes circular orbits, and in which the~objective is~to maximize per-shell capacity.
Optimizing for (e.g.) maximal temporal coverage or minimal operational collision risk would yield different counts.

The higher-altitude families accommodate more OSRs because they have larger orbital circumferences and wider eclipse-free LTAN bands.
Summing the~delivered-irradiance model from Appendix~\ref{app:reflected-spot} over these orbital rings (via Eq.~\ref{eq:altitude-shell-fluence-and-constellation-fluence}) using $120~\mathrm{m}\times120~\mathrm{m}$ reflectors gives year-averaged reflector-delivered surface flux (see Appendix~\ref{app:atmospheric-transmission}) equal to the~year-averaged natural surface flux from the~Sun (Figure~\ref{fig:multishell-power-to-base}).

\begin{table}[ht]
\centering
\caption{Representative geometric packing of the~Mars-base-warming constellation. Reflector totals are rounded to the~nearest 100. Operationally collision-safe constellations may require fewer total reflectors, relying on individually larger reflectors for equivalent power delivery.}
\label{tab:constellation-sizing-summary}
\begin{tabular}{lrrrrrrr}
\toprule
$K_{\rm ref}$ &
\makecell{Altitude\\range (km)} &
\makecell{Shell\\count} &
\makecell{Inclination\\range ($^\circ$)} &
\makecell{Rings per\\shell, $P$} &
\makecell{Reflectors per\\ring, $S$} &
\makecell{Total\\rings} &
\makecell{Total\\reflectors} \\
\midrule
12 & 508--623   & 24 & 93.22--93.56 & 3--6  & 58--82 & 110 & 7,800 \\
11 & 628--876   & 50 & 93.58--94.41 & 5--8  & 51--89 & 345 & 21,400 \\
10 & 881--1172  & 59 & 94.43--95.58 & 6--10 & 59--89 & 451 & 32,400 \\
9  & 1177--1432 & 52 & 95.60--96.78 & 7--13 & 47--93 & 482 & 32,300 \\
\midrule
Total & 508--1432 & 185 & 93.22--96.78 & 3--13 & 47--93 & 1,388 & 93,900 \\
\bottomrule
\end{tabular}
\end{table}

\end{document}